\documentclass[fleqn,useAMS,usenatbib]{mnras}

\usepackage[T1]{fontenc}
\usepackage{ae,aecompl}

\usepackage{pdflscape}
\usepackage{graphicx}	
\usepackage{amsmath}	
\usepackage{amssymb}	
\usepackage{xcolor}

\usepackage{txfonts}

\newcommand{\kms}{\hbox{${\rm km\;s}^{-1}$}}
\newcommand{\hi}{H~\textsc{i}}   

\newcommand{\Msun}{\ensuremath{M_{\sun}}}

\newcommand{\logmstar}{\ensuremath{\log \, (M_{\star}/M_{\sun})}}

\newcommand{\cbs}{Composite Bulges Survey}
\newcommand{\sfourg}{S\ensuremath{^{4}}G}
\newcommand{\atlas}{ATLAS\ensuremath{^{\rm3D}}}

\newcommand{\re}{\ensuremath{R_{e}}}
\newcommand{\rbrk}{\ensuremath{R_{\rm brk}}}

\newcommand{\hthree}{\ensuremath{h_{3}}}
\newcommand{\hfour}{\ensuremath{h_{4}}}
\newcommand{\Imfit}{\textsc{Imfit}}
\newcommand{\dpa}{\ensuremath{\Delta \mathrm{PA}}}
\newcommand{\dpamax}{\ensuremath{\Delta \mathrm{PA}_{\mathrm{max}}}}
\newcommand{\deltaaic}{\ensuremath{\Delta \mathrm{AIC}}}

\newcommand{\new}[1]{}
\defcitealias{rc3}{RC3}
\defcitealias{erwin20b}{Paper~I}

\title[Classical Bulges and Nuclear Discs in NGC~4608 and NGC~4643]{Composite Bulges -- II. Classical Bulges and Nuclear Discs in Barred Galaxies: The Contrasting Cases of NGC 4608 and NGC 4643}
\author[P. Erwin et al.]{
Peter Erwin,$^{1,2}$\thanks{E-mail: erwin@mpe.mpg.de}
Anil Seth,$^{3}$
Victor P. Debattista,$^{4}$
Marja Seidel,$^{5}$
Kianusch Mehrgan,$^{1,2}$
\newauthor
Jens Thomas,$^{1,2}$
Roberto Saglia,$^{1,2}$
Adriana de Lorenzo-C{\'a}ceres,$^{6,7}$
Witold Maciejewski,$^{8}$
\newauthor
Maximilian Fabricius,$^{1,2}$
Jairo M{\'e}ndez-Abreu,$^{6,7}$
Ulrich Hopp,$^{1,2}$
Matthias Kluge,$^{1,2}$
\newauthor
John E. Beckman,$^{6,7,9}$
Ralf Bender,$^{1,2}$
Niv Drory,$^{10}$
and Deanne Fisher$^{11}$ \\
$^{1}$Max-Planck-Insitut f\"{u}r extraterrestrische Physik, Giessenbachstrasse, 85748 Garching, Germany \\
$^{2}$Universit\"{a}ts-Sternwarte M\"{u}nchen, Scheinerstrasse 1, D-81679 M\"{u}nchen, Germany \\
$^{3}$Department of Physics and Astronomy, University of Utah, 115 South 1400 East, Salt Lake City, Utah 84112, USA \\
$^{4}$Jeremiah Horrocks Institute, University of Central Lancashire, Preston PR1 2HE, UK \\
$^{5}$IPAC, California Institute of Technology, 1200 East California Boulevard, Pasadena, CA 91125, USA \\
$^{6}$Instituto de Astrof\'{\i}sica de Canarias, C/ Via L\'{a}ctea s/n, 38200 La Laguna, Tenerife, Spain \\
$^{7}$Departamento de Astrof\'{\i}sica, Universidad de La Laguna, Avda. Astrof\'{\i}sico Fco. S\'{a}nchez s/n, 38200, La Laguna, Tenerife, Spain \\
$^{8}$ Astrophysics Research Institute, Liverpool John Moores University, Twelve Quays House, Egerton Wharf, Birkenhead CH41 1LD, UK \\
$^{9}$Consejo Superior de Investigaciones Cient\'ificas, Spain \\
$^{10}$McDonald Observatory, The University of Texas at Austin, 1 University Station, Austin, Texas 78712, USA \\
$^{11}$Center for Astrophysics and Supercomputing, Swinburne University of Technology, Hawthorn, Australia}

\date{Accepted XXX. Received YYY; in original form ZZZ}

\pubyear{2021}

\begin{document}
\label{firstpage}
\pagerange{\pageref{firstpage}--\pageref{lastpage}}
\maketitle

\begin{abstract} 

We present detailed morphological, photometric, and stellar-kinematic
analyses of the central regions of two massive, early-type barred
galaxies with nearly identical large-scale morphologies. Both have
large, strong bars with prominent inner photometric excesses that we
associate with boxy/peanut-shaped (B/P) bulges; the latter constitute
$\sim 30$\% of the galaxy light. Inside its B/P bulge, NGC 4608 has a
compact, almost circular structure (half-light radius $\re \approx 310$
pc, S\'ersic $n = 2.2$) we identify as a classical bulge, amounting to
12.1\% of the total light, along with a nuclear star cluster ($\re \sim
4$ pc). NGC 4643, in contrast, has a nuclear disc with an unusual
broken-exponential surface-brightness profile (13.2\% of the light), and
a very small spheroidal component ($\re \approx 35$ pc, $n = 1.6$; 0.5\%
of the light). IFU stellar kinematics support this picture, with NGC
4608's classical bulge slowly rotating and dominated by high velocity
dispersion, while NGC 4643's nuclear disc shows a drop to lower
dispersion, rapid rotation, $V$-\hthree{} anticorrelation, and elevated
\hfour. Both galaxies show at least some evidence for $V$-\hthree{}
\textit{correlation} in the bar (outside the respective classical bulge
and nuclear disc), in agreement with model predictions. Standard
2-component (bulge/disc) decompositions yield $B/T \sim 0.5$--0.7 (and
bulge $n > 2$) for both galaxies. This overestimates the true
``spheroid'' components by factors of four (NGC 4608) and over 100 (NGC
4643), illustrating the perils of naive bulge-disc decompositions
applied to massive barred galaxies.

\end{abstract}

\begin{keywords}
galaxies: structure -- galaxies: bulges -- galaxies: spiral
\end{keywords}

\section{Introduction} 

Disc galaxies have traditionally been said to have two main stellar
components: a flattened, rotationally supported \textit{disc} and
(optionally) a rounder, kinematically hot \textit{bulge}. The latter --
often characterized as akin to a small elliptical galaxy embedded
within the disc -- is thought to be especially prominent in the earliest
spirals and in lenticular/S0 galaxies. Older studies have suggested that
the bulge constitutes $\sim 40$--60\% of the total stellar light in such
systems \citep[e.g.,][]{simien86}, a result shared by some recent
decompositions of large, SDSS-based samples
\citep[e.g.,][]{oohama09,kim16a}.

Bulges are important for understanding galaxy evolution for a number of
reasons. First, their formation is supposed to involve early
mergers -- with the possibility of further growth via minor mergers --
so their presence and size provide potential clues about the early
stages of galaxy formation \citep[e.g.,][and references
therein]{brooks16}. In addition, a number of studies have shown that
star formation anticorrelates with the presence of bulges, which
suggests that the presence or growth of bulges is a possible
prerequisite for the cessation of star formation, or even a possible
\textit{cause} of it, a phenomenon sometimes termed ``morphological
quenching'' \citep[e.g.,][]{martig09,martig13,bluck14,lang14,eales20}.
Finally, there are the well-known correlations between bulge
characteristics -- particularly central velocity dispersions or bulge
luminosity/mass -- and supermassive black hole (SMBH) masses
\citep[e.g.,][and references therein]{kormendy13,saglia16}. In this
context, the idea that bulges are in effect smaller, lower-mass
elliptical galaxies embedded within larger discs seems to match up quite
well with the idea of SMBH-galaxy correlations that extend from massive
elliptical galaxies down to the small bulges of disc galaxies.

However, our understanding of how bulges form, grow, and correlate with
other galaxy properties has been complicated in the last couple of
decades by the realization that bulges are not all alike, and that not
all bulges are necessarily similar to small elliptical galaxies. This
began with the discovery that some bulges were unusually flattened or
dominated by rotation, or marked by the presence of clearly disclike
phenomena such as spiral arms or nuclear bars -- almost as if they were
more like small \textit{discs} embedded within the main disc
\citep[see][and references therein]{kormendy04}. 

A second development was the demonstration that some bulges in edge-on
galaxies had characteristics -- boxy or peanut shapes, cylindrical
stellar rotation, associated in-plane gas and stellar kinematics
indicative of bars -- indicating that even though they were clearly not
disclike structures, they \textit{were} the vertically thickened inner
parts of bars
\citep[e.g.,][]{kuijken95,merrifield99,bureau99a,lutticke00b,
chung-bureau04,bureau06}. Such structures are usually referred to as
``boxy/peanut-shaped'' (B/P) bulges; although they are thicker than
discs, in origin, structure, and kinematics they are emphatically
different from the traditional idea of a bulge.  More recent imaging
studies have shown that these structures can be identified in galaxies
with intermediate or low inclinations as well, appearing in the form of
``box + spurs'' or ``barlens'' morphologies inside bars
\citep{erwin-debattista13,laurikainen14,
athanassoula15,herrera-endoqui17,li17,laurikainen17,erwin-debattista17,
kruk19}; their frequency is a strong function of galaxy stellar mass,
becoming almost ubiquitous in the most massive barred galaxies
\citep{erwin-debattista17,li17}.

All such ``non-classical'' bulges have tended to be lumped together
under the term ``pseudobulges'', with the complementary term ``classical
bulges'' denoting the more traditional kinematically hot,
elliptical-like spheroids. Although there continues to be a tendency to
interpret these developments as meaning that the central regions of disc
galaxies are dichotomous -- that they can have \textit{either} a
classical bulge \textit{or} some (single) kind of pseudobulge -- there
has also been a growing awareness that ``bulges'' may sometimes be made
up of \textit{multiple, coexisting stellar components}. This was first
clearly articulated from a theoretical perspective by
\citet{athanassoula05}, who suggested that the centers of discs could
harbor potentially any combination of classical bulges, ``disc-like''
bulges, \textit{and} the B/P bulges of bars. Recent observational work
has clearly shown individual examples of galaxies with bulges that are
made up of at least two distinct components, either from a morphological
perspective based on profiles or image decompositions
\citep[e.g.,][]{laurikainen14,athanassoula15,lasker16}, or from
combinations of decompositions and analyses of the stellar kinematics
\citep[e.g.,][]{mendez-abreu14,erwin15a}.

Evidence for multiple types of bulges has been found for the Milky Way
itself: recent work has established that the dominant bulge in the Milky
Way is the B/P bulge of its bar
\citep[e.g.,][]{shen10,bland-hawthorn16,debattista17a}, with good
evidence for a low-mass nuclear disc
\citep[e.g.,][]{launhardt02,schonrich15,nogueras-lara19,gallego-cano20,
sormani20} -- but also little evidence for any significant classical
bulge \citep[e.g.,][]{bland-hawthorn16}. In addition, careful analysis
of M31 shows that \textit{its} bulge consists of two components: a
dominant B/P bulge belonging to its bar and a sub-dominant classical
bulge, with the former having about twice the stellar mass of the latter
\citep{athanassoula06,opitsch18,blana-diaz17,blana-diaz18}.

\begin{figure*}
\begin{center}
\hspace*{-3.0mm}\includegraphics[scale=0.92]{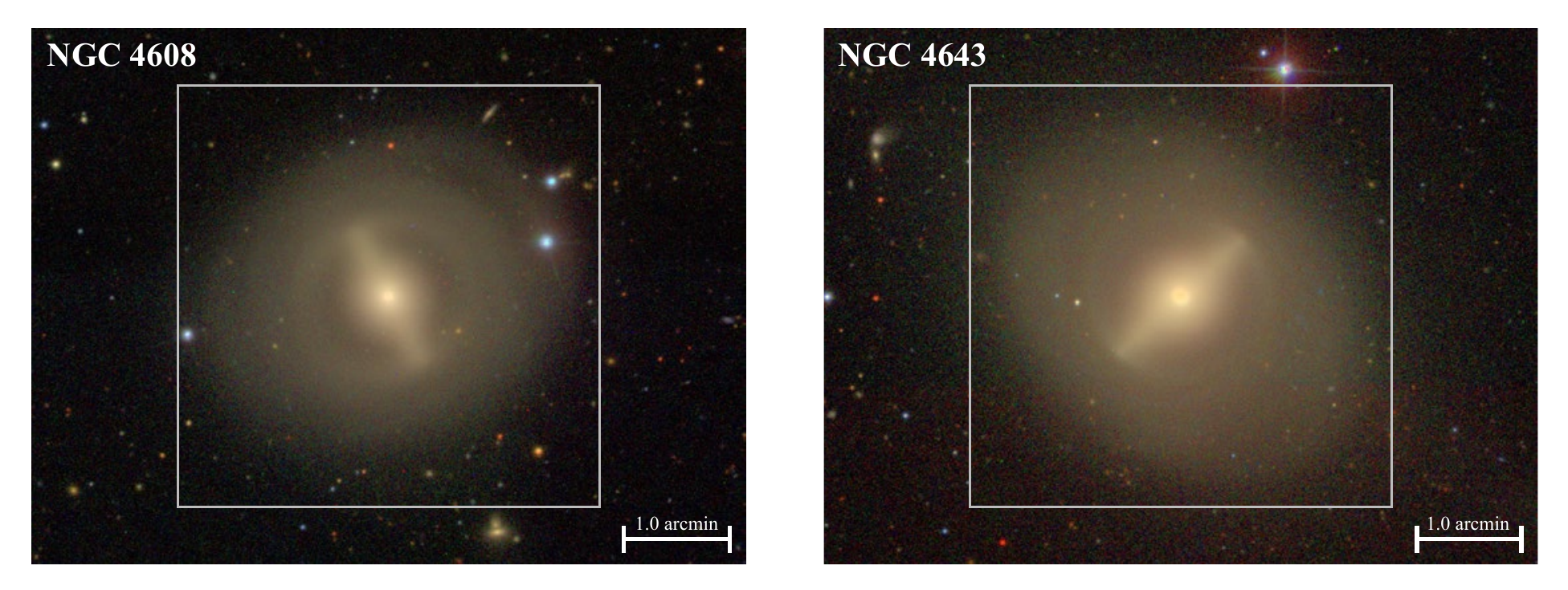}
\end{center}

\caption{SDSS $gri$ colour-composite images (David W.\ Hogg, Michael R.\ Blanton, and the Sloan
Digital Sky Survey Collaboration) for the two galaxies studied
in this paper. The light grey boxes outline the
large-scale regions shown (using \textit{Spitzer} IRAC1 isophotes) in
the leftmost panels of Figures~\ref{fig:n4608-overview} and
\ref{fig:n4643-overview}.}\label{fig:n4608-n4643-color}

\end{figure*}

The question of what ``bulges'' in disc galaxies actually are is now a
critical one. In particular, we would like to know how often disc galaxies
have spheroidal, kinematically hot classical bulges, how often they have
nuclear discs or discy pseudobulges, and how often they have B/P bulges.
We would also like to know how often, and in which ways, these different
structures can coexist in the same galaxy, and what fraction of a
galaxy's stars can be found in the different inner components.

Identifying, enumerating, and understanding these distinct forms of
bulges is important, because they have very different formation
mechanisms, and potentially different effects on their host galaxies. As
noted above, classical bulges are supposed to be the result of violent,
gas-rich mergers very early in galaxy's history. Discy pseudobulges
(nuclear discs) are thought to form from bar-driven gas inflow and
subsequent star formation, and are thus the result of internal
(``secular'') processes that are unrelated to mergers, and which require
the prior formation of a bar
\citep[e.g.,][]{kormendy04,wozniak09,cole14, baba20}. B/P
bulges also require the formation of a bar, since they form out of (and
remain part of) the bar; they are also the result of internal processes,
albeit ones rather different from gas inflow
\citep[e.g.,][]{combes81,
combes90,raha91,quillen02,martinez-valpuesta04,debattista05,debattista06,
quillen14,sellwood20}. All of this is clearly relevant to more
general bulge--host-galaxy correlations and processes. For example, it
is easy to see how classical bulges might correlate with SMBHs, while a
connection with nuclear-disc formation is less obvious, and the
formation of B/P bulges would at first glance seem quite
unrelated to SMBH growth.\footnote{\citet{fragkoudi16} do
suggest that bars with B/P bulges could be less efficient (than bars
without B/P bulges) at promoting gas inflow to the central kpc, leading
to lower nuclear fueling rates.}

To address these issues, we have undertaken a multi-wavelength imaging
and spectroscopic observing campaign -- the \cbs{} \citep[][hereafter
``Paper~I'']{erwin20b} -- aimed at identifying and characterizing
\textit{all} the different components that may reside in the centers of
massive disc galaxies, from bars -- and their boxy/peanut-shaped inner
components -- down to nuclear discs (and nuclear rings and nuclear bars)
and even to nuclear star clusters \citep[NSCs;][]{neumayer20}. This
study is based around a mass- and volume-limited sample of approximately
fifty nearby, early- and intermediate-type (S0--Sbc) disc galaxies, each
observed in the optical and near-IR with the \textit{Hubble Space
Telescope} (\textit{HST}), and with planned observations of all galaxies
using the MUSE integral field spectrograph.

In this paper we present a detailed study of inner structures of two
galaxies from this sample. This paper both serves as a model for
analysis of the rest of our sample (using both high-resolution
near-infrared imaging and 2D stellar kinematics), and enables a
comparison of how our detailed approach contrasts with the simpler
approaches frequently used in galaxy decompositions, especially for
large samples. The galaxies we study here are early-type disc galaxies
with similar masses and morphologies, seen at very similar orientations:
NGC 4608 (SB0, $\logmstar = 10.38$) and NGC~4643 (SB0/a, $\logmstar =
10.79$); see Figure~\ref{fig:n4608-n4643-color} for large-scale optical
views of both.\footnote{See Section~\ref{sec:overview} for the source of
their classifications and stellar masses.} Both galaxies have
inclinations $i \sim 35$--40\degr{} and are strongly barred, with the
bar oriented close to the minor axis of its parent galaxy. In a general
morphological sense, they are close to being twins, and there is
evidence that both galaxies' bars contain B/P bulges, making them even
more similar.\footnote{Evidence for B/P bulges can be found in
\citet{laurikainen14}, \citet{athanassoula15}, \citet{laurikainen17},
and Section~\ref{sec:particulars}.} And yet, as we will show, their
inner regions are quite different: one galaxy (NGC~4608) hosts a
classical bulge (and probably a nuclear star cluster) inside its B/P
bulge, while the other (NGC~4643) has a massive nuclear
disc\footnote{Previously suggested by \citet{erwin-sparke03} and
\citet{erwin04}.} with an unusual broken-exponential surface-brightness
profile. Inside this nuclear disc is a smaller spheroidal structure that
is probably too large to be a nuclear star cluster, but is about an
order of magnitude smaller and less massive than the first galaxy's
classical bulge.

Section~\ref{sec:data} of this paper discusses the parent sample and the
data sources for the two galaxies, while Section~\ref{sec:overview}
summarizes their general characteristics. Section~\ref{sec:decomp} is
devoted to our detailed, 2D decompositions of near-IR images of both
galaxies, while Section~\ref{sec:kinematics} uses published and archival
2D stellar-kinematic data to test and validate the morphological results
of the preceding decompositions. Section~\ref{sec:discussion} includes
discussion of the nuclear disc in NGC~4643 and the implications of our
analysis for simplistic, two-component decompositions of large galaxy
samples; Section~\ref{sec:summary} summarizes our findings.

\section{Parent Sample and Data Sources}\label{sec:data} 

A fuller description of our sample and observing strategy is
presented in \citetalias{erwin20b}; here, we provide a brief overview.

The \cbs{} is designed to probe the morphology and stellar kinematics
and populations of the bulge regions (e.g., the inner 1--2 kpc) of
massive disc galaxies. It is based on a volume- and mass-limited sample
with an upper limit on distances of 20 Mpc, so that with \textit{HST}
imaging we can resolve down to $\sim 15$ pc or better in the near-IR
(assuming a PSF FWHM $\approx 0.15\arcsec$ in the F160W band).

The sample is defined so as to include all S0--Sbc galaxies in
\citet[][RC3]{rc3} with Galactic latitude $|b| > 20\degr$, distances
$\leq 20$ Mpc, stellar masses $\geq 10^{10} \Msun$, and inclinations
between $35\degr$ and $60\degr$. Since we plan to augment our dataset
with archival and future spectroscopy from the Very Large Telescope
(VLT), we limited the positions on the sky to $\delta \leq 20\degr$. The
result was a total of 54 galaxies spanning a range of environments from
the local field to the Virgo Cluster. One galaxy (NGC~5363) was
subsequently identified as a probable elliptical rather than an S0, so
we are left with a total of 53 disc galaxies.

\subsection{\textit{HST} Data from the Composite Bulges Survey} 

The primary dataset is a consistent set of optical and near-IR images
obtained with \textit{HST}, using the Wide Field Camera 3 (WFC3) in both
its UVIS and IR modes (Cycle 25, Proposal ID 15133). Since we are first
and foremost interested in the underlying stellar structure, with as
little confusion due to dust as possible, the main emphasis of the
proposal was on full-field WFC3-IR imaging using the F160W filter, with
a total integration time of 600s divided into four dithered exposures.
For the sake of efficiency in readout time and data storage, the optical
imaging was restricted to the C1K1C aperture, which is a $1024 \times
1024$-pixel subset of the full WFC3-UVIS array; we obtained four
dithered exposures in each of the F475W and F814W filters, for total
integration times of 700s in F475W and 500s in F814W.\footnote{See
\citetalias{erwin20b} for notes on slight variations in this scheme for
six of the galaxies, which did not apply to the galaxies studied in this
paper.} This enabled us to obtain all exposures for a given galaxy
within a single \textit{HST} orbit.

Individual exposures in each band were combined using the Python-based
Drizzlepac code. After some experimentation, we adopted output image
scales of 0.03\arcsec{} pixel$^{-1}$ for the optical images and
0.06\arcsec{} pixel$^{-1}$ for the F160W images, with \texttt{pixfrac} =
0.7. Since the galaxies are larger than the WFC3-IR field of view, it is
impossible to accurately estimate the sky background from the \textit{HST}
images; accordingly, we turned sky subtraction off during the
processing. (We account for the sky background as part of our modeling
process; see Section~\ref{sec:decomp}.)

For purposes of photometric calibration, we started with the standard
calibration to the Vega-magnitude (VEGAMAG) F160W system in
the F160W filter. We converted these to 2MASS $H$-band equivalents
following the suggestion of \citet{riess11}, which uses the 2MASS $J - H$
colours. For red galaxies like NGC~4608 and NGC~4643, the correction works out to
$m_H \approx m_{\rm 160w,vega} + 0.025$.

\subsection{Other Imaging Data}\label{sec:other-imaging-data} 

For large-scale decompositions, we make use of \textit{Spitzer} IRAC1
(3.6\micron) images of NGC~4608 and NGC~4643. The image for NGC~4643 comes from
the \textit{Spitzer} Survey of Stellar Structure in Galaxies
\citep[\sfourg;][]{sheth10}, with the latter's final mosaic pixel scale of
0.75 \arcsec/pixel. NGC~4608 was not part of \sfourg{} so we use the
archive-generated mosaic image (Program ID 10043, PI Kartik Sheth), with
the default archive mosaic pixel scale of 0.6 \arcsec/pixel.\footnote{The
observation of NGC~4608 is part of an extension to \sfourg{} meant to
fill in the missing elliptical and S0 galaxies
\citep{sheth13,knapen14}.} For both galaxies, we estimated the residual
sky background as the mean of the median values of $\sim 50$--100 $20
\times 20$-pixel boxes located well away from the galaxies and from
bright stars.

For the purpose of estimating how both galaxies would fare under
bulge-disc decompositions at redshifts typical of large, SDSS-based
studies (Section~\ref{sec:bt-ratios}), we also used $r$-band images from Data
Release 7 of SDSS \citep{york00,abazajian09}.

\subsection{Integral Field Unit Spectroscopic Data: SAURON and MUSE} 

Two-dimensional stellar-kinematic information, obtained with the SAURON
instrument, is available for the central regions of both galaxies from
\atlas{} \citep{cappellari11a-atlas3d-1,krajnovic11a-atlas3d-2}. SAURON
\citep{bacon01} is an integral field spectrograph composed of
0.94\arcsec{} lenslets with a $33\arcsec \times 41\arcsec$ field of
view, which spans the inner $3 \times 3$ kpc of each galaxy. We used the
stellar-kinematics tables (containing the Gauss-Hermite $V$, $\sigma$,
\hthree{} and \hfour{} parameters) from the main \atlas{} web
site\footnote{\url{http://www-astro.physics.ox.ac.uk/atlas3d/tables}
}. The instrumental resolution $\sigma_{\rm inst}$ is $\approx 105$
\kms, which is comfortably below the central velocity dispersions of
both galaxies (130 \kms{} for NGC~4608 and 147 \kms{} for NGC~4643,
according to HyperLEDA).

We additionally made use of SAURON observations from \citet{seidel15b} for
NGC~4643. These consist of seven separate pointings which together cover
the whole of the bar region ($\sim 100 \times 80\arcsec$) rather than
just the single-pointing \atlas{} observation, which only covers the
innermost region.

NGC~4643 has been observed with the VLT's MUSE instrument as part of the
TIMER project \citep{gadotti19}. MUSE \citep{bacon10} is an integral
field spectrograph using image slicers to sample a $1\arcmin \times 1\arcmin$
field of view (in the Wide Field Mode) at 0.2\arcsec{} per spaxel
(spatial pixel). The spectrograph covers a wavelength range of
4800--9300 \AA, with a resolution that ranges from $R =
\lambda/\Delta\lambda = 1770$ to 3590 over that same range and a
spectral sampling of 1.25 \AA/pixel.

Preliminary MUSE stellar and gas kinematics, along with a
stellar-population analysis, have been presented for NGC~4643 in
\citet{gadotti19,gadotti20} and \citet{bittner20}, which we refer to
later. We also performed our own stellar-kinematic analysis of the
publicly available datacube (ESO Phase 3 reduced datacube, part of the
``DEEP MUSE'' datastream). We inspected the datacube for possible
sky-subtraction errors, but did not find any; in any case, our interest
is in the bright central region of NGC~4643, dominated by the galaxy
light. To do this, we used the penalized, pixel-fitting pPXF code
\citep{cappellari04-ppxf,cappellari17} to derive stellar kinematics in
the usual form of a parameterized line-of-sight velocity distribution
(LOSVD) -- mean stellar velocity $V$, velocity dispersion $\sigma$, and
the Gauss-Hermite \hthree{} and \hfour{} parameters -- using the MILES
stellar-template library \citep{sanchez-blazquez06,falcon-barroso11}. We
first performed a fit to the combined spectrum of the full datacube in
order to identify a subset of template spectra with non-zero weights.
Using the latter as a restricted template library, we then fit spectra
(using the wavelength range 4800--7400 \AA) from all the individual
spaxels, masking out narrow regions around emission lines. This differed
somewhat  from the analysis of \citet{gadotti19,gadotti20}, who used
Voronoi-binned spectra, fitted wavelength ranges of 4750--5500 \AA{}
\citep{gadotti19} or 4800--8950 \AA{} \citep{gadotti20}, and the E-MILES
model library of SSP template spectra \citep{vazdekis15}, but our
results are almost identical to theirs.

We also performed a kinematic analysis using our own non-parametric
pixel-fitting code (\citealt{mehrgan19}, J. Thomas et al.\ 2020, in
preparation), in part to verify that the results were not dependent on
the particulars of which code was used; further details of this analysis
will be presented elsewhere.

\begin{figure*}
\begin{center}
\hspace*{-3mm}\includegraphics[scale=0.87]{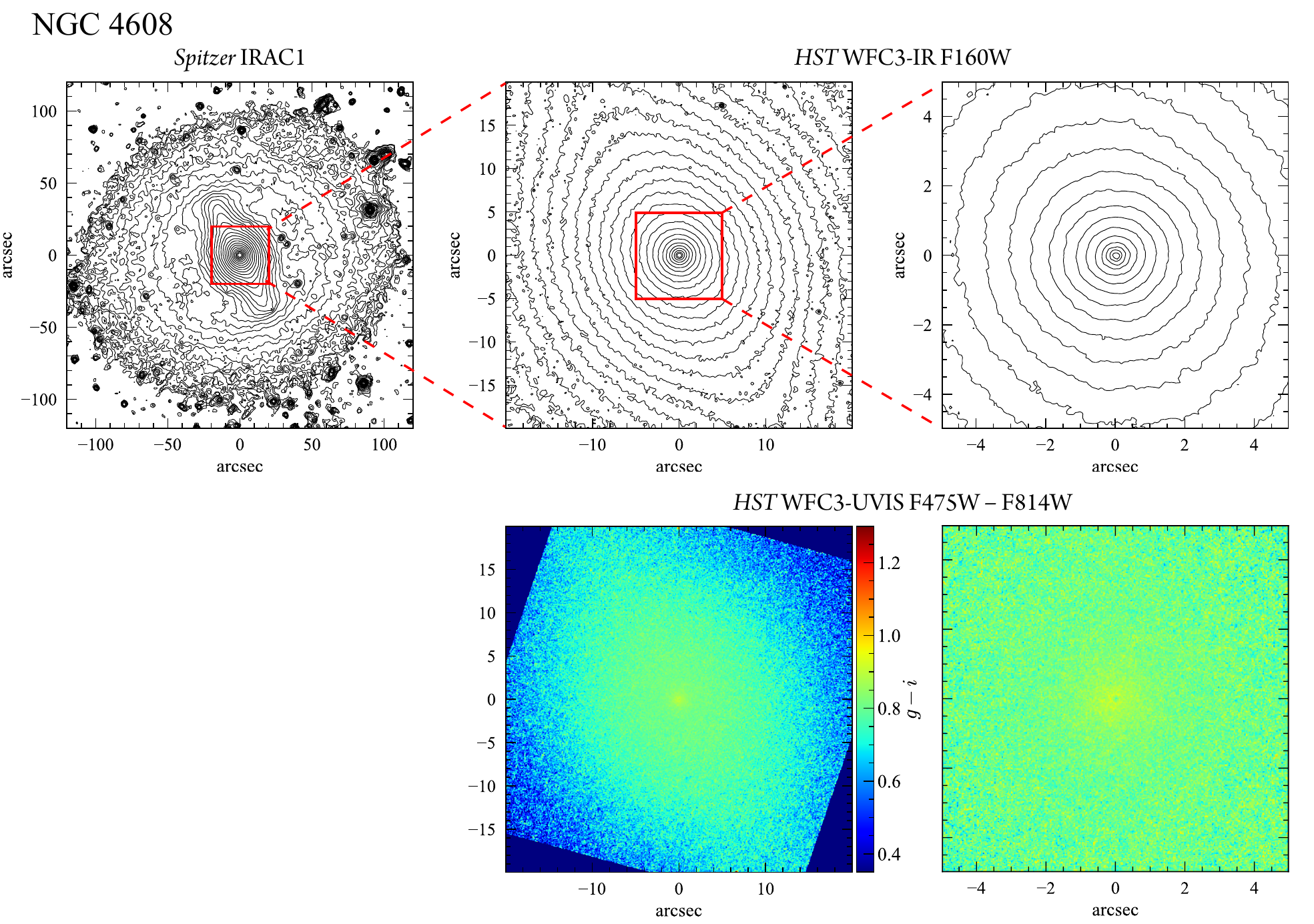}
\end{center}

\caption{Imaging overview of NGC~4608. Top row: from left to right,
logarithmically spaced \textit{Spitzer} IRAC1 isophotes (left,
median-smoothed with width = 5 pixels, contour spacing of 0.1 des) and
two close-ups showing \textit{HST} WFC3-IR F160W isophotes (middle,
median-smoothed with same width; right, unsmoothed). Bottom row:
\textit{HST} WFC3-UVIS colour maps (F475W $-$ F814W, converted to $g -
i$) on the same spatial scales and smoothing as the corresponding F160W
isophote plots. (Unless otherwise explicitly noted, all figures in this
paper have standard astronomical orientation: N = up, E =
left)}\label{fig:n4608-overview}

\end{figure*}

\begin{figure*}
\begin{center}
\hspace*{-3mm}\includegraphics[scale=0.87]{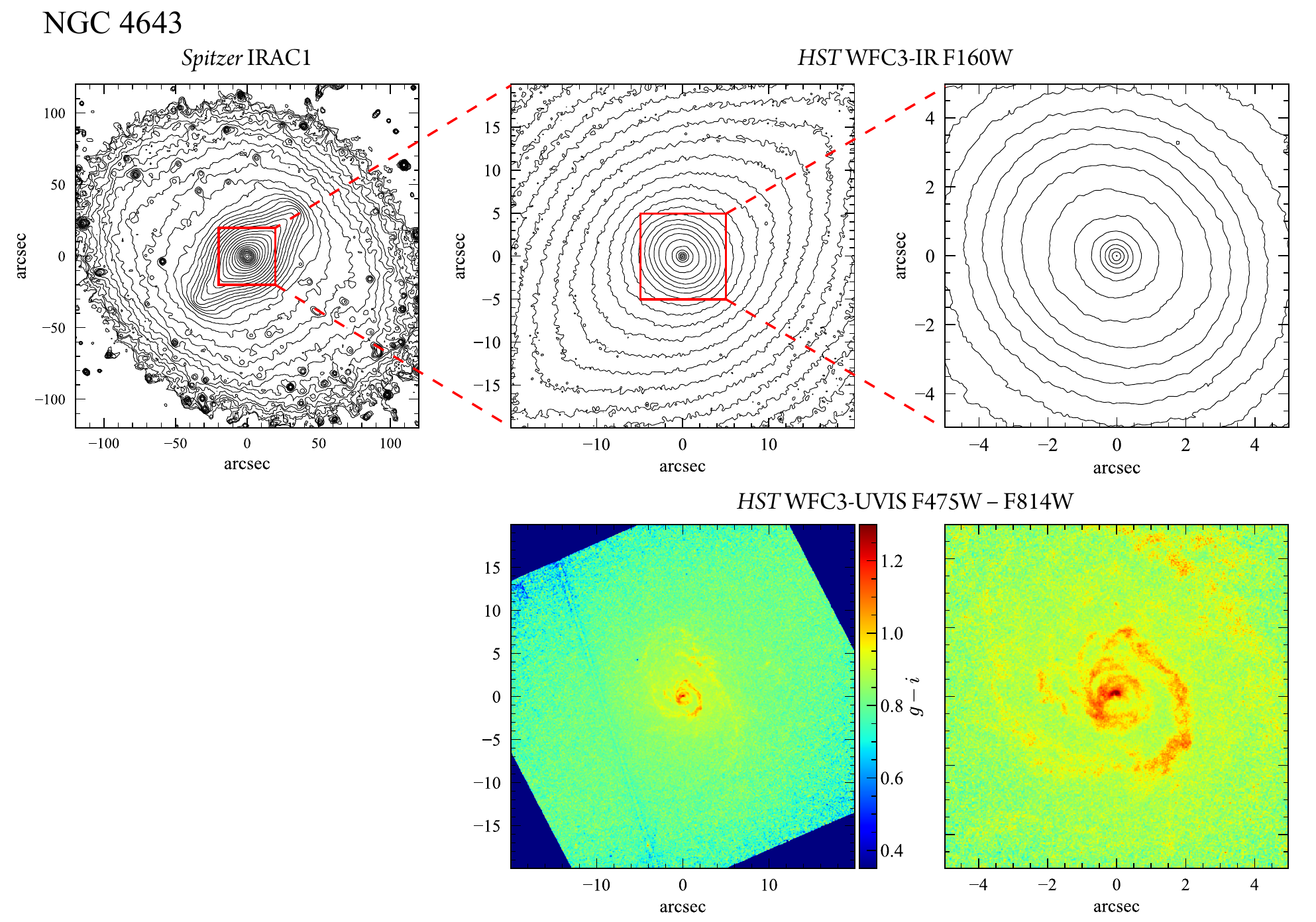}
\end{center}

\caption{Same as Figure~\ref{fig:n4608-overview}, but now showing
NGC~4643.}\label{fig:n4643-overview}

\end{figure*}

\section{An Overview of NGC 4608 and NGC 4643}\label{sec:overview}

NGC~4608 and NGC~4643 are both massive, strongly barred, early-type disc
galaxies. The former is located in the Virgo Cluster at a distance of
$17.3 \pm 0.8$ Mpc \citep[surface-brightness-fluctuation
distance;][]{cantiello18}; the latter is in the nearby field
(approximately 2.4 Mpc away from NGC~4608), with a redshift-based
distance estimate of $19.3 \pm 2.9$ Mpc.\footnote{Assuming $H_{0} = 72$
\kms~kpc$^{-1}$ and a Virgocentric-infall-corrected redshift of 1392
\kms{} from HyperLEDA, along with an assumed uncertainty of 15\%.}
These distances give scales of 83.9 pc/\arcsec{} for NGC~4608 and
93.6 pc/\arcsec{} for NGC~4643.

NGC~4608 is classified in \citetalias{rc3} as SB0 and NGC~4643 as SB0/a;
very faint, wispy spiral arms account for the latter classification
\citep[e.g.,][]{erwin-sparke03}. They are very similar in overall
orientation as well, with inclinations of 36\degr{} and 38\degr{},
respectively; in both cases their bars are positioned almost perfectly
along each galaxy's minor axis. The position angles of their discs are
100\degr{} and 53\degr{} for NGC~4608 and NGC~4643, respectively. (Disc
orientations are based on the isophote shapes of their outer discs and
are taken from \citealt{erwin05b} and \citealt{erwin08}.)

Their stellar masses are also similar. For both galaxies, we estimated
$H$-band $M/L$ ratios from their global $g - i$ colours, using the
relation of \citet{roediger15}. The colours are based on the SDSS Model
$g$ and $i$ magnitudes from Data Release 6, while the $H$-band
magnitudes are from the 2MASS Extended Source Catalog (all values
retrieved from NED, corrected for Galactic extinction using the
\citealt{schlafly11} values). This gives us $\logmstar = 10.381$ for
NGC~4608 and 10.789 for NGC 4643, so the latter galaxy is about 2.5 times
more massive than the former.

They thus present themselves (aside from the factor of two difference in
stellar mass) as very much alike, as noted by, e.g., the
\textit{Carnegie Atlas} \citep{carnegie}: ``NGC 4643 is nearly identical
to NGC 4608 \ldots{} The description there applies to NGC
4643 as well.''

The bars and inner (``bulge'') regions appear at first glance to be very
similar as well. Both bars show strong, narrow isophotes which
transition from being sharply pointed (``discy'') to having rectangular
(``boxy'') ends as one moves out in radius. The inner isophotes are
significantly rounder, which encourages the idea (mostly erroneous, as
we will argue) that we are seeing dominant classical bulges in the
centers of both galaxies.

Nonetheless, there are hints of differences when the inner isophotes are
examined more carefully. In particular, the isophotes of NGC~4643 switch
from the bar orientation (PA $\approx 135\degr$) to a mildly elliptical
shape nearly perpendicular to the bar -- and closely aligned with that
of the outer disc: PA $\approx 52\degr$ for $r \la 5\arcsec$. While this
is in principle consistent with an oblate classical bulge, previous
analysis, including unsharp masking, of ground-based images has found
evidence for a nuclear disc with a possible stellar nuclear ring in this region
\citep{erwin-sparke03,erwin04}. \citet{gadotti19} have used MUSE data
(see Section~\ref{sec:n4643-muse}) to show that the stellar kinematics
in this region is dominated by rotation, arguing for the presence of a
nuclear disc.

\begin{figure}
\begin{center}
\hspace*{-3mm}\includegraphics[scale=0.95]{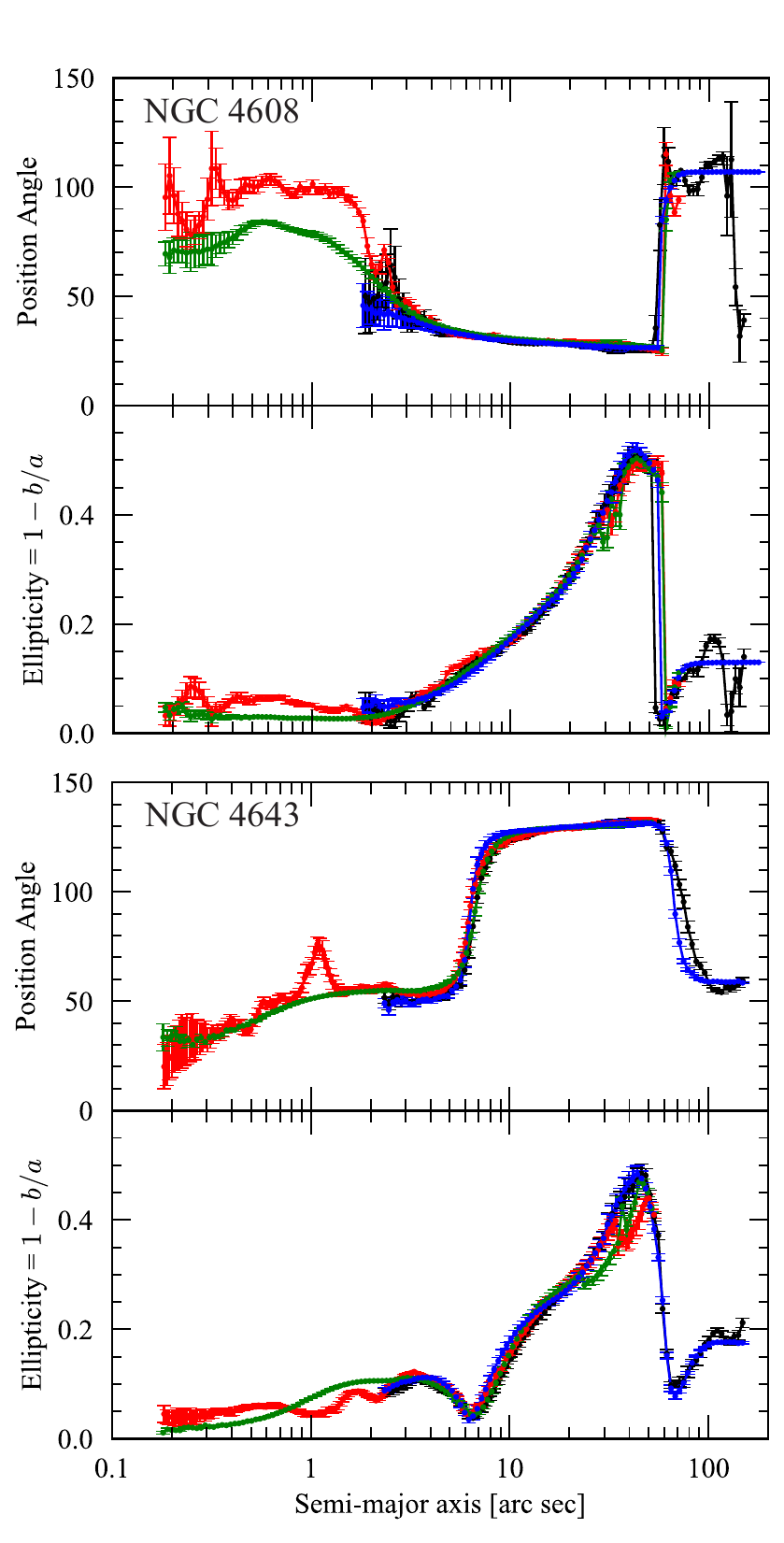}
\end{center}

\caption{Ellipse fits to \textit{Spitzer} IRAC1 images (black) and
\textit{HST} F160W images (red) of both galaxies (top: NGC~4608; bottom:
NGC~4643), along with ellipse fits to the best-fitting model images
(blue = model fit to IRAC1 image, green = model fit to F160W image).
}\label{fig:efits}

\end{figure}

Figures~\ref{fig:n4608-overview} and \ref{fig:n4643-overview} show
large-scale (IRAC1) and close-up views (F160W) of the near-IR isophotes
for both galaxies, along with \textit{HST} colourmaps using the F475W and
F814W WFC3-UVIS images. Figure~\ref{fig:efits} shows the position angle
and ellipticity of ellipses fitted to the IRAC1 and F160W isophotes
using the \textsc{iraf} \texttt{ellipse} package. The colourmaps show
that NGC~4608 has no evidence for dust or recent star formation, just a
slight inward reddening trend. NGC~4643, on the other hand, shows spiral
dust lanes in the inner $r \la 10\arcsec$. The latter are not strong
enough to cause significant distortions in the F160W image, though they
may be responsible for the small variations in ellipticity and position
angle at $a \la 2\arcsec$ visible in the ellipse fits for this galaxy.

\section{Morphological Decompositions}\label{sec:decomp}

\subsection{General Outline}

Our general goal is to dissect each galaxy into astrophysically distinct
stellar components, which might have different dynamics and/or formation
mechanisms. We do this by modeling each image as the sum of multiple
2D image functions, with the functions chosen to represent
plausible galaxy components (discs, bars, rings, nuclear star clusters,
etc.). We put particular emphasis on the accurate modeling of the inner
regions (roughly, the inner 1--2 kpc), in order to identify distinct
``bulge'' components and sub-components.

We combine modeling of low-resolution, whole-galaxy images from
\textit{Spitzer} with the high-resolution, small-field-of-view WFC3-IR
F160W images from our \textit{HST} observations. This emphasis on
near-IR images stems from our interest in the stellar structure,
undistorted (as much as possible) by dust extinction. Although NGC~4608
appears to be dust-free (e.g., Figure~\ref{fig:n4608-overview}),
NGC~4643 has some dust lanes in the inner $\sim 1$ kpc. These are not
strong enough to noticeably distort the F160W image, but would make
fitting the optical images problematic
(Figure~\ref{fig:n4643-overview}). We note that we are implicitly
assuming no strong colour gradients involving the F160W and IRAC1 bands,
so that we can treat them as describing the same stellar structure.

Our fits for each galaxy are done in a two-stage process. First, we
focus on fitting the large-scale, sky-subtracted IRAC1 image using disc-
and bar-related components; the innermost components (e.g., bulge,
nuclear disc, nuclear star cluster) are treated approximately, since the
IRAC1 images lack the spatial resolution to properly sample these parts
of the galaxy. When an acceptable fit has been achieved, we switch to
fitting the F160W image, where we can refine and better constrain the
innermost components of the model. The F160W modeling begins with the
best-fitting model and parameter values from the IRAC1 fits, translating
size and orientation parameters appropriately. Since the galaxies are
much larger than the F160W field of view, we fix the outermost component
(the main disc) to the best-fit values from the IRAC1 fits, with the
exception of intensity parameters (i.e., $I_{0}$ for the
BrokenExponential component we use for the main disc in both galaxies),
which are left as free parameters of the F160W fit. We also included a
constant-background (``FlatSky'') term to account for the sky background
in the F160W image (since the galaxies are larger than the WFC3-IR field
of view, the background cannot be determined directly).

For each galaxy, we start with very simple models, such as bulge +
disc (i.e., an elliptical S\'ersic component + an elliptical Exponential
component). We inspect the residuals to find where they are particularly
bad (data much brighter than model or vice-versa) and to look for
possible shapes suggestive of missing components. We then gradually add
additional components to the model, using basic morphological analysis
and prior studies as a guide. For example, since both NGC~4608 and
NGC~4643 are strongly barred, including bar components in the models is
a logical step. Since NGC~4643 has previously been identified as having
a possible nuclear stellar ring \citep[e.g.,][]{erwin04} and/or a
nuclear disc \citep[e.g.,][]{gadotti19}, we also test both of these
as possible additional components. The specific sequence of models for
NGC~4608 is discussed in Section~\ref{sec:n4608-fit-example}; the process
for NGC~4643 is described in Section~\ref{sec:n4643-progress}.

The ``goodness'' of these successive fits is evaluated in a relative fashion:
by comparing how well a new model reduces systematic, patterned
residuals produced by previous models; by judging how well a new model
reproduces certain features in the isophotes; and by looking for
significant improvement in the Akaike Information Criterion (AIC,
\citealt{akaike74}). The AIC is useful because, unlike the reduced
$\chi^{2}$, it can be used to compare different models fit to the same
data. Most discussions of model comparison argue that
$\deltaaic < -10$ indicate a clearly better fit for the model with the
lower AIC value. However, from our experience in fitting models to
galaxy images, we find that most additions to a model, even relatively
trivial ones, can usually produce large negative values of \deltaaic,
and we only consider improvements of $\deltaaic < -1000$ to be truly
useful.

Fits were done with the \textsc{Imfit} package 
\citep{erwin15b}\footnote{\url{https://www.mpe.mpg.de/~erwin/code/imfit}},
using standard $\chi^{2}$ minimization and the (default)
Levenberg-Marquardt minimizer; we used the data values to estimate the
per-pixel uncertainties under the usual Gaussian approximation of
Poisson statistics. For the IRAC1 images, pixel values were converted
back to ADUs, with an assumed A/D gain of 3.7 used in the fit to convert
ADUs to detected photons. For the F160W images, our reduction generated
final images with detected photons (``electrons'') as the final pixel
value, so no further processing was necessary. Because the images
include foreground stars, background galaxies, and occasional image
defects, we prepared masks based on running SExtractor \citep{bertin96}
to identify individual sources, with detected sources transformed into
circles (for stars) or ellipses (for background galaxies) with radii or
semi-major axes equal to small multiples of the SExtractor
\texttt{A\_IMAGE} parameter.

All fits include convolution with the appropriate PSF image. For the
IRAC1 images, we used the official in-flight Pixel-Response-Function
images\footnote{\url{https://irsa.ipac.caltech.edu/data/SPITZER/docs/
irac/calibrationfiles/psfprf/}}, downsampled to the appropriate pixel
from the PRF scale of 0.24\arcsec/pixel. We used the PRF image at
column,row = 129,129 as an approximation to the central location
of the galaxy in the image; since our final modeling of the central
regions of each galaxy is based on the \textit{HST} images, more
accuracy than this for the IRAC1 PSF is not needed. For the F160W
images, we generated appropriate PSF images using the \texttt{grizli}
software,\footnote{\url{https://github.com/gbrammer/grizli}} which
inserts the ``empirical PSF'' images of
\citet{anderson16-wfc3ir-psfs}\footnote{\url{http://www.stsci.edu/~
jayander/STDPSFs/WFC3IR/}} into the four individual F160W exposures and
then runs them through the same drizzling process we use to prepare our
final F160W images, extracting the final PSF image from the combined,
drizzled image \citep{mowla19}.

\subsection{Determining Best-Fit Models: The Example of NGC~4608}\label{sec:n4608-example}

As an example of our successive modeling approach, consider
Figure~\ref{fig:n4608-residuals-progress}, which shows residuals from fitting
successively more complex models to the IRAC1 and F160W images of NGC~4608. 

The top row of the figure shows the results of our fits to the IRAC1
image. Panel~b shows a simplistic reference (``bulge+disc'')
decomposition, where we use only an Exponential (disc) and a S\'ersic
(bulge) component. The residuals indicate this is a terrible fit, both
due to the lack of a bar component and because the disc is not a simple
exponential. Panel~c shows the next stage, where we add our new,
two-component bar model (Appendix~\ref{app:bar}) \textit{and} change the
disc component from a simple exponential to a broken-exponential. This
is a dramatic improvement, with much-reduced residuals and a much lower
AIC ($\deltaaic \sim -3.5 \times 10^{5}$). In the next stage (panel~c to
panel~d), we use the strong, ringlike residual surrounding the bar in
panel~c, along with the clear evidence from the isophotes (and previous
classifications of this galaxy) for an inner ring, to motivate adding a
GaussianRing component to the model. As panel~d shows, this clearly
reduces the ringlike residual \textit{and} the deficits immediately
inside the ring, though the residuals in the center and the disc outside
the ring are unchanged; the improvement in AIC ($\deltaaic \approx
-13500$) is significant, but less than the previous stage. Finally, in
panel~e we show what happens when we replace the uniform GaussianRing
component with one that has an azimuthally varying surface brightness
(the GaussianRingAz component, described in Appendix~\ref{app:ring}).
This is formally a meaningful improvement ($\deltaaic \approx -1700$),
though the residuals appear almost unchanged (except for the partial
spiral residuals immediately north and south of the east and west ends
of the bar, respectively). Comparison of the isophotes of the data and
model images (e.g., upper and middle panels of
Figure~\ref{fig:n4608-bestfit-irac1}) shows that this change
\textit{does} make the model match the data better in the inner-ring
region (e.g., the fact that the inner ring visibly decays as the angle
with respect to the bar major axis increases). Nonetheless, at this
stage we seem to be reaching the limits of significant improvements for
modeling the IRAC1 image.

We then shift to modeling the F160W image, starting with the final
IRAC1 model. Since most of the main disc is outside the field of view of
the F160W image, we re-use the BrokenExponential component by holding
its parameters fixed to their best-fit values from the IRAC1 fit, with
the exception of the central surface brightness, which was left as a
free parameter. We also include a uniform-background component
(\Imfit's FlatSky component) to account for the (unknown) sky background
in the F160W image. The resulting fit is generally excellent, except
for clear nuclear excess in data (panel~g; also suggested by nuclear
excess in higher-resolution WFC3-UVIS optical images). The addition of a
compact Gaussian to represent a possible nuclear star cluster 
clearly improves the fit and removes the nuclear excess from the
residuals (panel~h). Since the central isophotes are essentially
circular, and since this component is only partially resolved, we fix
its shape to be circular.

\subsection{Estimating Parameter Uncertainties}

Determining the uncertainty of fitted parameters for fits such as these
is problematic. Although \Imfit{} provides nominal error estimates when
run with the (default) Levenberg-Marquardt minimization algorithm, we
elected to use the bootstrap-resampling option, which produced slightly
more generous (and probably accurate) results, including asymmetric
uncertainties; these are reported in our best-fit parameter tables
(Tables~\ref{tab:n4608-decomp} and \ref{tab:n4643-decomp}). Bootstrap
resampling also allows for plotting uncertainty distributions between
parameters, which allows the identification of correlated uncertainties.
Samples of such correlations are presented in
Appendix~\ref{app:bootstrap}.

As inspection of the tables will show, these error estimates are in
general extremely small -- typically $< 1$\%. This is probably because
the data we work with is very high $S/N$, and the error estimates assume
that the model is a correct representation of the data, with the
uncertainty coming only from random, per-pixel errors (e.g, from Poisson
processes and Gaussian readout noise). As an example of how misleading
this can be, consider the simple bulge/disc (S\'ersic + exponential)
decomposition of the IRAC1 image of NGC~4608 (panel~b of
Figure~\ref{fig:n4608-residuals-progress}). The best-fit position-angle
and ellipticity of the disc component are $111.4\degr \pm 0.2\degr$ and
$0.209 \pm 0.003$, respectively -- while the best-fit values for the
(BrokenExponential) disc in our final IRAC1 model are $106.6\degr \pm
0.2\degr$ and $0.127 \pm 0.001$. Clearly, the formal errors of the fit
do not adequately represent uncertainties in what we might consider
general, model-independent properties of the galaxy or its main
subcomponents. (A similar point was made by \citealt{gao17} regarding
the dependence of ``bulge'' properties on the inclusion or exclusion
of other components in their 2D models of galaxies.)

We investigated the possible influence of uncertainties in the sky
background subtraction by fitting versions of the IRAC1 images
that had been perturbed with the $\pm 1$-$\sigma$ sky uncertainties
(the latter determined by 1000 rounds of bootstrap resampling
applied to the sky-level determination method). These produced
variations in the fitted parameters roughly equivalent to, or smaller
than, the bootstrap-based uncertainties.

\subsection{Some Particulars: Broken-Exponential Discs and a Two-Component
Bar Model}\label{sec:particulars} 

Preliminary fitting experiments with the IRAC1 images indicated that the
best results were obtained when we treated the main disc as having a
\textit{broken-exponential} radial surface-brightness profile
\citep[e.g.,][]{erwin08,erwin15b}, where an inner, shallow exponential
zone breaks to a steeper outer exponential profile. This is handled with
the BrokenExponential component in \textsc{Imfit}, which combines
elliptical isophotes with a broken-exponential radial profile. We note that in
\citet{erwin08} neither galaxy shows a clear ``Type~II'' profile (though
NGC~4608 does in fact have a weak Type~II.i profile). However, those
profile classifications were based on the azimuthally averaged profile
\textit{outside} the bar region, while our interest here is in 2D
modeling of the entire galaxy, including the bar and its interior.
The necessity for a broken-exponential (i.e., Type II) profile for
2D fits to images of NGC~4608 was previously pointed out by \citet{gadotti08}.

\citet{laurikainen14} suggested, based on 2D fits to a ground-based
$K$-band image, that NGC~4643 possessed a ``barlens'', which is their
term for the B/P bulge seen at low inclinations. They fit the bar with a
``thin bar'' component (a 2D projection of a Ferrers ellipsoid) and an
exponential-like component for the B/P bulge, along with an extra
sub-exponential (S\'ersic $n = 0.7$) component for the ``bulge''. For
NGC~4608, \citet{laurikainen05} noted ``a large oval inside the bar, and
apparently a small spherical bulge'' in their $K$-band image.
\citet{athanassoula15} modeled the same image using a separate
``barlens'' component for the B/P bulge and a projected Ferrers
ellipsoid for the outer, thin part of the bar. Both galaxies are listed
as ``good examples'' of a thin bar plus a barlens in
\citet{laurikainen17}. We note that both galaxies are also massive
enough for the presence of B/P bulges to be statistically likely: using
the logistic-regression analysis of \citet{erwin-debattista17}, which
models the probability that a barred galaxy has a B/P bulge as a
function of the galaxy's stellar mass, the probabilities are 0.52 for
NGC~4608 and 0.85 for NGC~4643.

Given this evidence, we work from the assumption that the bars of these
two galaxies \textit{do} have B/P bulges, and that the latter are best
considered as components with rounder isophotes and steeper
surface-brightness profiles, \textit{added to} the more elongated,
shallow-surface-brightness-profile components of the outer (vertically
thin) part of the bar.

Our general logic is thus similar to that of \citet{laurikainen14} and
\citet{athanassoula15}, with the bar represented as the sum of two
components. (A very similar approach was followed by \citealt{neumann19}
for some of their barred galaxies.) We differ from them in using a new
component for the outer part of the bar, designed to better represent
(than is possible with S\'ersic or Ferrers functions) the
broken-exponential major-axis profile and the transition from discy to
boxy isophotes near the end of the bar. This new component
(``FlatBar''), is described in Appendix~\ref{app:bar}.

Unlike \citet{athanassoula15}, we put more emphasis on modeling the main
disc of the galaxy as well, in order to better constrain the modeling of
the bar region; this includes (as noted above) treating the main discs
of both galaxies as having broken-exponential surface-brightness
profiles; it also means using a separate component to model the
prominent inner ring surrounding the bar in NGC~4608. This allows us to
fit most or all of the parameters freely, in contrast to having to fix
such parameters as the bar length and the B/P bulge ellipticity by hand,
as Athanassoula et al.\ did.

\subsection{Fitting NGC 4608: Summary}\label{sec:n4608-fit-example}

\begin{figure*}
\begin{center}
\hspace*{-3mm}\includegraphics[scale=0.61]{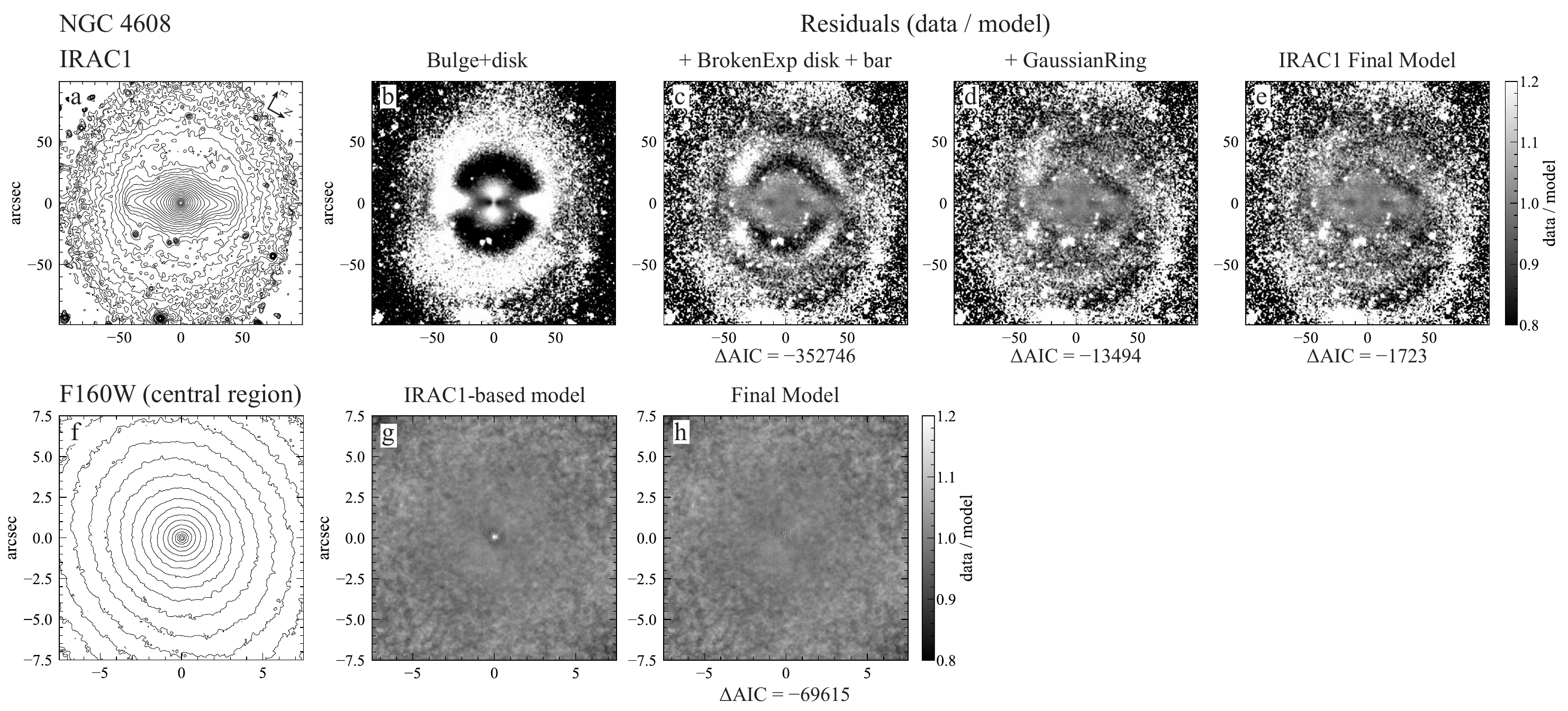}
\end{center}

\caption{Progressively more complex and accurate 2D image modeling of
NGC~4608. \textbf{Upper row:} logarithmically spaced, median-smoothed
isophotes of the fitted region of the \textit{Spitzer} IRAC1 image
(a) and ``residual-ratio'' images (data / model) from models fit to
this image (b--e). These are, from left to right: simple bulge/disc (S\'ersic +
exponential) model (b); addition of our two-component bar model
(Appendix~\ref{app:bar}) and replacement of exponential (disc) with 
BrokenExponential component (c); addition of GaussianRing component to
model the inner ring surrounding the bar (d); and replacement of the
GaussianRing with a GaussianRingAz (Appendix~\ref{app:ring}) component (e).
Numbers below each image show the relative improvement of the fit (change
in Akaike Information Criterion [AIC] relative to the previous fit). Note
that image orientation in this row is as observed.
\textbf{Lower row:} Isophotes from the inner region of
the \textit{HST} WFC3-IR F160W image (f) and residual-ratio images
from fits to the (full) F160W image. From left to right, these are: the ``final
model'' from the IRAC1 fits (g) and addition of a central Gaussian component to
represent a nuclear star
cluster (h). Note that in this row, image orientation is standard (N = up, E = left).
}\label{fig:n4608-residuals-progress}

\end{figure*}

\begin{figure*}
\begin{center}
\hspace*{-3mm}\includegraphics[scale=1.1]{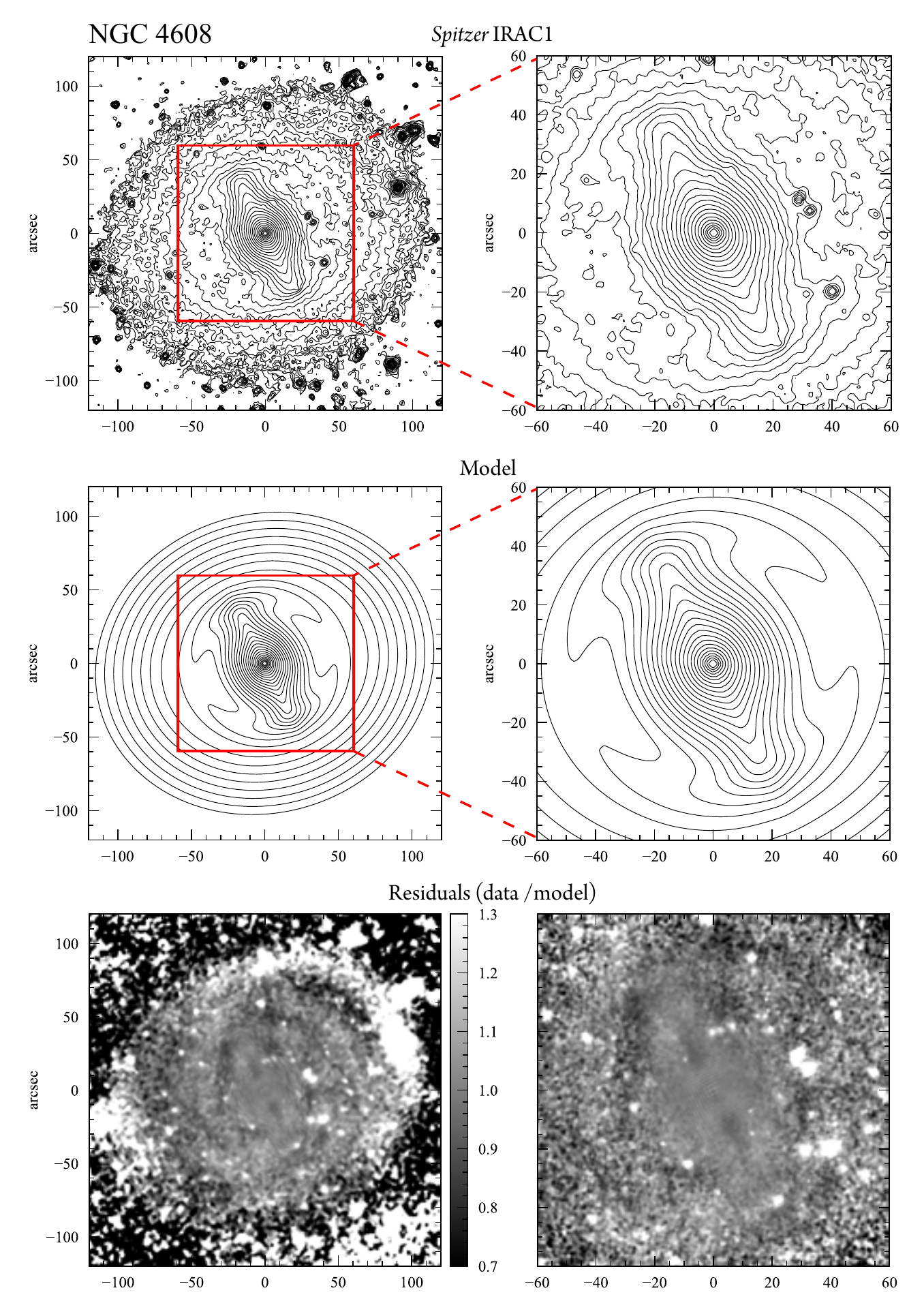}
\end{center}

\caption{Best-fitting model for NGC~4608, compared with \textit{Spitzer}
IRAC1 image. Upper and middle panels show logarithmically spaced
isophotes from IRAC1 image (top) and best-fitting model (middle); bottom
panels show residuals (data values divided by model values). The
large-scale images (left) were smoothed using a $5 \times 5$-pixel-wide
median filter.}\label{fig:n4608-bestfit-irac1}

\end{figure*}

\begin{figure*}
\begin{center}
\hspace*{-3mm}\includegraphics[scale=1.1]{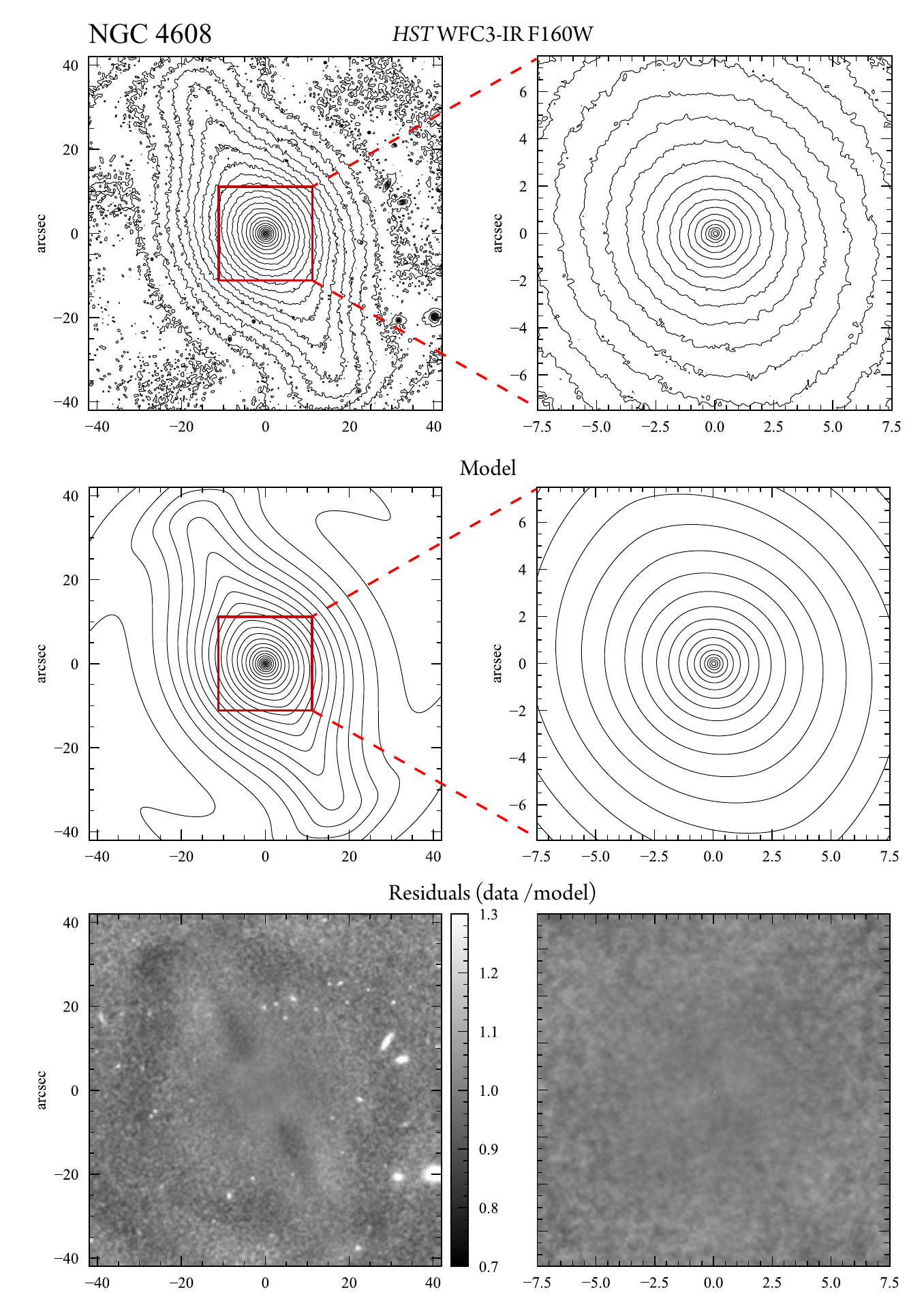}
\end{center}

\caption{Best-fitting model for NGC~4608, compared with \textit{HST}
WFC3-IR F160W image. Upper and middle panels show logarithmically spaced
isophotes from F160W image (top) and best-fitting model (middle); bottom
panels show residuals (data values divided by model values). The
large-scale images (left) were smoothed using a $9 \times 9$-pixel-wide
median filter.}\label{fig:n4608-bestfit-f160w}

\end{figure*}

(See Section~\ref{sec:n4608-example} and
Figure~\ref{fig:n4608-residuals-progress} for how we proceeded in
constructing and successively refining our best-fit model for this galaxy.)

Our final model for NGC~4608 uses the combination of \Imfit's
BrokenExponential component and the new GaussianRingAz component
(Appendix~\ref{app:ring}) for the disc and the inner ring surrounding
the bar; the new FlatBar component for the outer part of the bar and a
mildly elliptical Sersic\_GenEllipse component for the inner, B/P part
of the bar; a central round S\'ersic component; and a compact, circular
Gaussian. Table~\ref{tab:n4608-decomp} summarizes the best-fitting
parameter values, while Table~\ref{tab:fluxes} shows the luminosities
and relative fluxes of the different components, and
Figures~\ref{fig:n4608-bestfit-irac1} and \ref{fig:n4608-bestfit-f160w}
compare this model with the IRAC1 and F160W data. In addition, the upper
half of Figure~\ref{fig:efits} compares the ellipticity and
position-angle profiles of the best-fit model with those of the data.
This shows generally excellent agreement, except possibly for the inner
$r < 2\arcsec$, where the model isophotes are slightly too round and
differ in PA by $\sim 20\degr$.

The main/outer disc (the combination of the BrokenExponential and
GaussianRingAz components) accounts for 45\% of the total light; most of
this is in the BrokenExponential, with the ring component being only
3.3\%. The inner part of the BrokenExponential is effectively constant
(the fits converge to the upper limit we set on the parameter value);
this changes to an exponential scale length of $\approx 27\arcsec$ (2.16
kpc) beyond the break radius ($\rbrk = 70\arcsec$, 5.58 kpc). This outer
scale length is very similar to the value of 29\arcsec{} reported by
\citet{erwin08} for the disc outside the bar, based on an azimuthally
averaged $r$-band profile from an SDSS image.

The bar is represented by the combination of an inner S\'ersic component
(with very weakly boxy isophote shapes) for the B/P bulge and a FlatBar
component for the outer part of the bar; the latter has major-axis inner
and outer scale lengths of 50.8\arcsec{} and 3.7\arcsec, respectively,
and a break radius of $\sim 41\arcsec$, which is very similar to the
maximum-ellipticity length of $44\arcsec$ reported for the bar by
\citet{erwin05b}. The B/P-bulge part of the bar has an essentially
exponential surface-brightness profile ($n = 0.96$); it is slightly
misaligned with respect to the FlatBar component ($\Delta$PA $\approx
7\degr$, with the B/P bulge oriented slightly closer to the galaxy major
axis); this is consistent with the effects of projection operating on a
bar close to, but not perfectly aligned with, the galaxy minor axis (see
\citealt{erwin-debattista13} for a discussion of how projection effects
produce misalignments between the projected B/P bulge and the ``spurs''
of the outer part of the bar). Together, the two components are 43\% of
the total galaxy light, only slightly less than the main disc; the B/P
bulge by itself is 29\% of the light.

The inner, round (ellipticity $\approx 0.04$) S\'ersic component has $n
\approx 2.2$ and $\re \approx 3.7\arcsec = 310$ pc. Given that it is
significantly rounder than the disc (ellipticity $\approx 0.18$), we
view this component as a candidate classical bulge; it amounts to
$\approx 12.1$\% of the total galaxy light. The Gaussian component
inside this is a plausible nuclear star cluster, especially considering
its compact size ($\re \approx 4$ pc); its luminosity is only $\sim
0.056$\% of the total galaxy light. This combination of size and stellar
mass ($\sim 1.3 \times 10^{7} \Msun$) makes it reasonably typical for a
nuclear star cluster \citep[e.g., Fig.~7 in][]{neumayer20}.

We note that including this last (NSC) component has a moderate effect
on the derived parameters of the round, inner S\'ersic component (which
we identify as a potential classical bulge). Without the NSC component,
the latter has $n = 2.55$ and $\re = 4.14\arcsec$ (347
pc), both of which are about $\sim 13$--15\% larger than their values
when the NSC component is included. There are minor changes to the B/P
bulge component, at the level of $\sim 2$\% for $n$ and \re.

\begin{table}
\caption{NGC 4608: 2D Decomposition}
\label{tab:n4608-decomp}
\begin{tabular}{@{}llrl}
\hline
Component          & Parameter & Value & Units \\
\hline

 Gaussian           & PA & --- & \degr{} \\
 (NSC)  & $\epsilon$ & $0^{1}$ &  \\
                    & $\mu_{0}$ & $12.27 \; [+0.14, -0.23]$ & SB \\
                    & $\sigma$ & $0.058 \; [+0.004, -0.006]$ & \arcsec{} \\
                    &  & $4.8 \; [+0.4, -0.5]$ & pc \\
 Sersic             & PA & $82.9 \pm 0.5$ & \degr{} \\
 (bulge)            & $\epsilon$ & $0.0359 \pm 0.0005$ &  \\
                    & $n$ & $2.214 \; [+0.007, -0.008]$ &  \\
                    & $\mu_{e}$ & $16.514 \; [+0.006, -0.007]$ & SB \\
                    & $r_{e}$ & $3.67 \pm 0.02$ & \arcsec{} \\
                    &  & $308.0 \pm 1.5$ & pc \\
 Sersic\_GenEll     & PA & $33.00 \; [+0.07, -0.06]$ & \degr{} \\
 (bar: BP bulge)    & $\epsilon$ & $0.1129 \pm 0.0004$ &  \\
                    & $c_{0}$ & $0.099 \pm 0.002$ &  \\
                    & $n$ & $0.960 \pm 0.002$ &  \\
                    & $\mu_{e}$ & $17.348 \pm 0.003$ & SB \\
                    & $r_{e}$ & $10.342 \pm 0.008$ & \arcsec{} \\
                    &  & $867.4 \pm 0.7$ & pc \\
 FlatBar            & PA & $26.520 \pm 0.007$ & \degr{} \\
 (bar: outer)       & $\epsilon$ & $0.8708 \; [+0.0003, -0.0004]$ &  \\
                    & $\Delta$PA$_{\rm m}$ & $40.0 \pm 0.3$ &  \\
                    & $\mu_{0}$ & $17.831 \; [+0.002, -0.003]$ & SB \\
                    & $h_{1}$ & $46.3 \pm 0.2$ & \arcsec{} \\
                    &  & $3886 \; [+13, -14]$ & pc \\
                    & $h_{2}$ & $3.75 \pm 0.02$ & \arcsec{} \\
                    &  & $315 \pm 1$ & pc \\
                    & $R_{\rm brk}$ & $40.7 \pm 0.02$ & \arcsec{} \\
                    &  & $3412 \pm 2$ & pc \\
                    & $\alpha$ & $0.498 \pm 0.002$ & \arcsec$^{-1}$ \\
 GaussianRingAz     & PA & $106.2 \pm 0.2$ & \degr{} \\
 (inner ring)       & $\epsilon$ & $0.0850 \pm 0.0005$ &  \\
                    & $A_{\rm maj}$ & $21.667 \; [+0.006, -0.005]$ & SB \\
                    & $A_{\rm min,rel}$ & $2.26 \pm 0.02$ &  \\
                    & $R_{\rm ring}$ & $47.99 \pm 0.02$ & \arcsec{} \\
                    &  & $4025 \pm 1$ & pc \\
                    & $\sigma$ & $6.37 \pm 0.03$ & \arcsec{} \\
                    &  & $534 \pm 2$ & pc \\
 BrokenExp          & PA & $106.6  \pm 0.2$ & \degr{} \\
 (main disc)        & $\epsilon$ & $0.127 \pm 0.001$ &  \\
                    & $\mu_{0}$ & $20.366 \pm 0.004$ & SB \\
                    & $h_{1}$ & $6000^{1}$ & \arcsec{} \\
                    &  & $50300$ & pc \\
                    & $h_{2}$ & $27.0 \; [+0.1, -0.2]$ & \arcsec{} \\
                    &  & $2262\; [+12, -19]$ & pc \\
                    & $R_{\rm brk}$ & $69.7 \; [+0.2, -0.3]$ & \arcsec{} \\
                    &  & $5847 \; [+19, -24]$ & pc \\
                    & $\alpha$ & $0.32 \; [+0.02, -0.03]$ & \arcsec$^{-1}$ \\
\hline 
\end{tabular}

\medskip

Summary of the final 2D decomposition of NGC~4608. Column 1: \Imfit{}
component names. Column 2: Parameter names. Column 3: Best-fit parameter
value; note that for size parameters, we include sizes in arcsec and
also in pc. Errors are nominal 68\% confidence intervals from bootstrap
resampling, and should be considered underestimates; errors for linear
sizes do \textit{not} include uncertainties in the distance. Column 4:
Units of the parameter (``SB'' = $H$-band mag arcsec$^{-2}$). For all
parameters except the final BrokenExponential (``main disc''), values come from fitting
the \textit{HST} WFC3-IR F160W image; for the latter, values come from
fitting the \textit{Spitzer} IRAC1 image. Surface-brightness values are
for $\mu_{H}$. Notes: 1 = at limit of parameter-value boundaries;
uncertainty undefined.

\end{table}

\subsection{Fitting NGC 4643}\label{sec:n4643-fit}

\subsubsection{Determining the Best-Fit Model}\label{sec:n4643-progress}

\begin{figure*}
\begin{center}
\hspace*{-3mm}\includegraphics[scale=0.61]{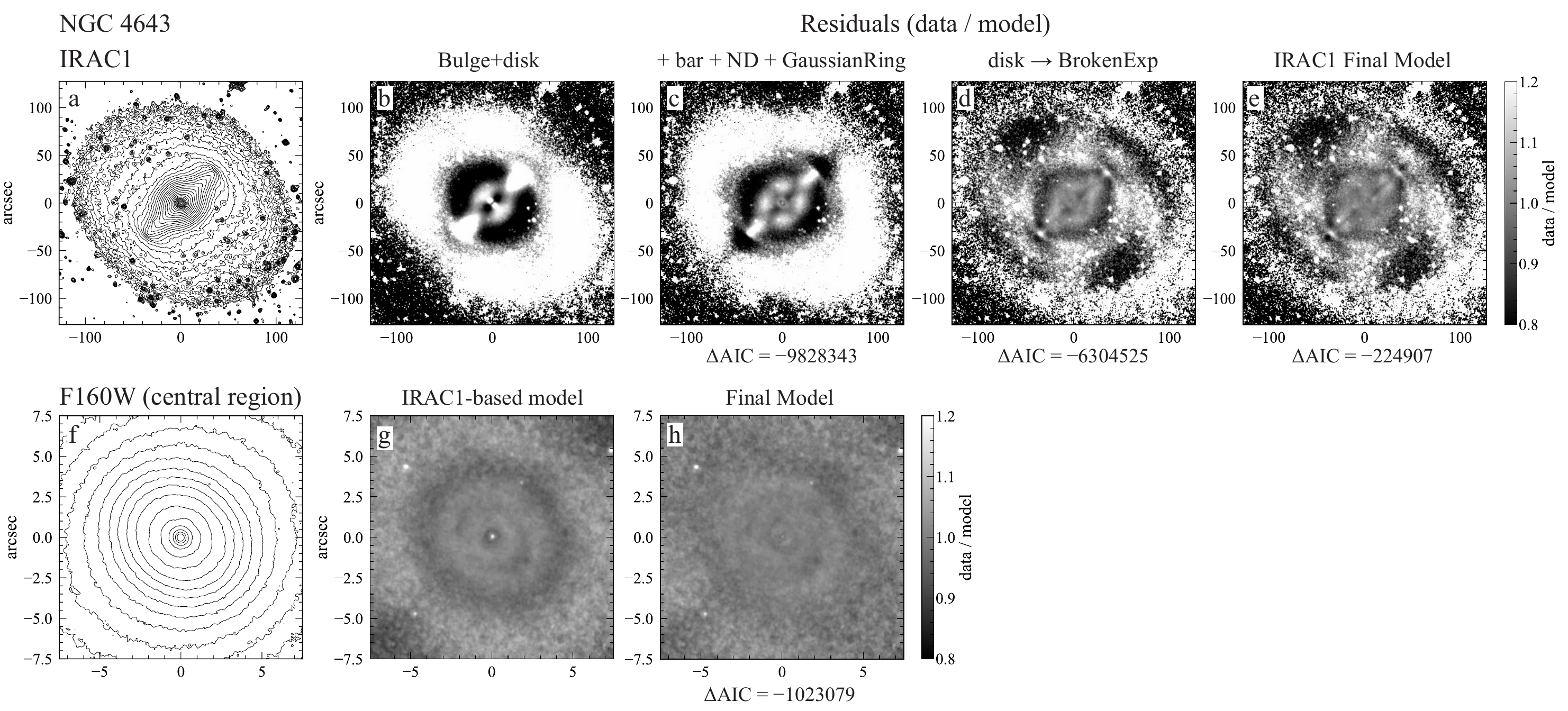}
\end{center}

\caption{As for Figure~\ref{fig:n4608-residuals-progress}, but now
showing progressive modeling of NGC~4643. \textbf{Upper row:}
logarithmically spaced, median-smoothed isophotes of the fitted region
of the \textit{Spitzer} IRAC1 image (a) and ``residual-ratio'' images
(data / model) from models fit to the image (b--e). These are (from left
to right): simple bulge/disc (S\'ersic + exponential) model (b);
addition of bar + nuclear disc + nuclear ring (c); replacement of outer
exponential with BrokenExponential (d); replacement of exponential
B/P-bulge component with S\'ersic component (e). \textbf{Lower row:}
Isophotes from the \textit{HST} WFC3-IR F160W image (f) and
residual-ratio images from fits to the (full) image. From left to right, these
are: the ``final model'' from the IRAC1 fits (g) and replacement of inner
exponential + GaussianRing with BrokenExponential + Sersic components to
better model nuclear disc and compact classical bulge (or NSC) (h).}\label{fig:n4643-residuals-progress}

\end{figure*}

In Figure~\ref{fig:n4643-residuals-progress} we show progressively more
complex fits to the images of NGC~4643. Panel~b shows a basic bulge+disc
decomposition, which -- as we saw for NGC~4608 -- is completely
inadequate. Panel~c shows a more complicated
fit with our two-component bar model (using, for simplicity, an
Exponential for the B/P bulge rather than a S\'ersic function)
\textit{and} a first attempt at a nuclear disc model, consisting of an
elliptical exponential and a GaussianRing component (the latter for the
putative nuclear ring). This is a significantly better fit (e.g,
$\deltaaic \approx -9.8 \times 10^{6}$), but the residuals are
still strong and systematic. 

Panel~d shows what is almost the same model, except that we have
replaced the outer Exponential with a BrokenExponential component. This
brings a dramatic improvement in the residuals (and a further
$\deltaaic \approx -6.3 \times 10^{6}$), which illustrates the
importance of getting the main disc component right. (Note that the fit
in the bar region has improved as well.) Finally, panel~e shows what
happens when we replace the B/P-bulge exponential component with one
having a S\'ersic profile. This is an improved fit, with the
best-fitting S\'ersic index for the B/P-bulge component $n \approx
0.52$.

In panel~g we show this ``best'' model from the IRAC1 image fits
(panel~e) fit to the F160W image. The residuals are generally small,
but there are some troublesome systematics in the inner $r \la
10\arcsec$, with alternating radial deficits and excesses. In addition,
we found that the GaussianRing component in this fit converged to a very small
radius ($\sim 0.06\arcsec$) and a very broad width ($\sigma \sim
2.5\arcsec$). This disagrees with the apparent size of the ring ($r \sim
3.0\arcsec$) and indicates that the Exponential + GaussianRing model for
the center of this galaxy is not a good match to the data. Careful
inspection of the F160W image (panel~f) and profile cuts through the
center hinted that the morphology might be better represented by the
combination of a broken-exponential structure for the nuclear disc and a
compact, round component at $r \la 0.5\arcsec$. Replacing the inner
Exponential + GaussianRing with a BrokenExponential component and a
central, round S\'ersic component produced our final model (panel~h),
with clearly better residuals in the central region and $\deltaaic \sim
-1.0 \times 10^{6}$. There remain some very faint systematic patterns in
the residuals at this stage, perhaps indicative of very subtle ring or
spiral morphology within the nuclear disc.

\begin{figure*}
\begin{center}
\hspace*{-3mm}\includegraphics[scale=1.1]{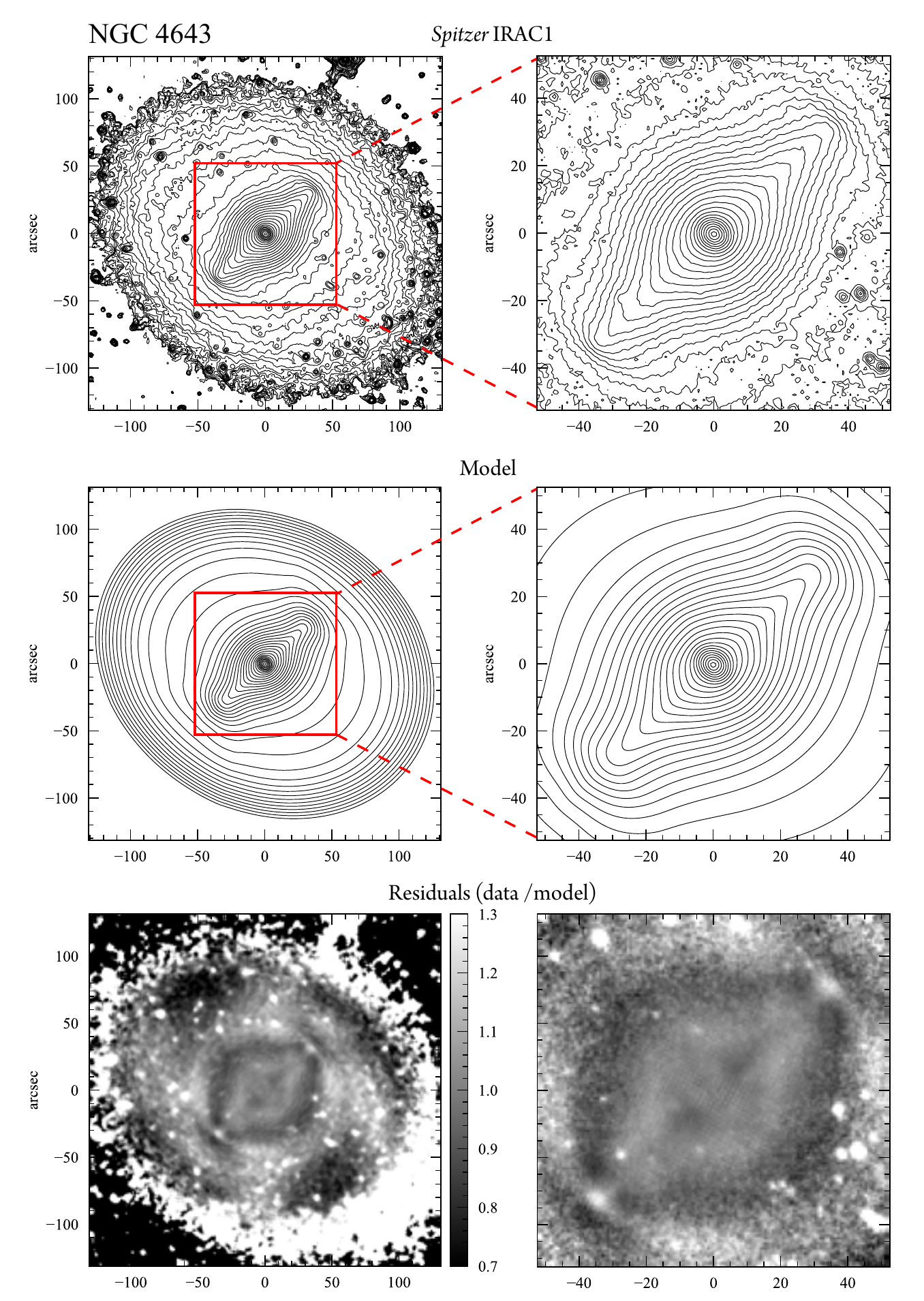}
\end{center}

\caption{Best-fitting model for NGC~4643, compared with \textit{Spitzer}
IRAC1 image. Upper and middle panels show logarithmically spaced
isophotes from IRAC1 image (top) and best-fitting model (middle); bottom panels
show residuals (data values divided by model values). 
The large-scale images (left) were smoothed using a $5 \times 5$-pixel-wide
median filter.}\label{fig:n4643-bestfit-irac1}

\end{figure*}

\begin{figure*}
\begin{center}
\hspace*{-3mm}\includegraphics[scale=1.1]{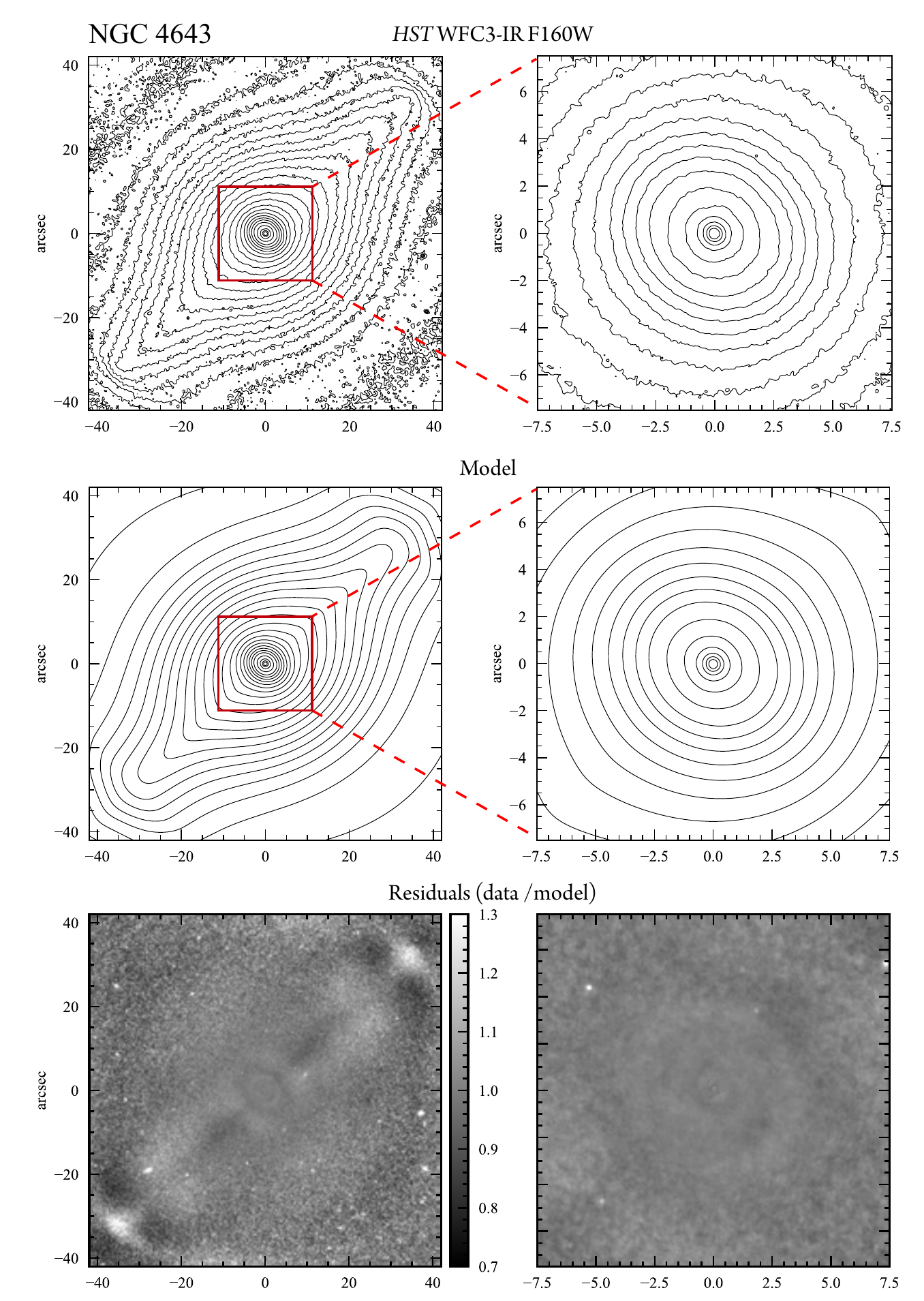}
\end{center}

\caption{Best-fitting model for NGC~4643, compared with \textit{HST}
WFC3-IR F160W image. Upper and middle panels show logarithmically spaced
isophotes from F160W image (top) and best-fitting model (middle); bottom panels
show residuals (data values divided by model values). The
large-scale images (left) were smoothed using a $9 \times 9$-pixel-wide
median filter.}\label{fig:n4643-bestfit-f160w}

\end{figure*}

\subsubsection{Summary}

Our final model for NGC~4643 uses \Imfit's BrokenExponential component
for the main disc; the new FlatBar component for the outer part of the
bar and a mildly elliptical Sersic\_GenEllipse component for the inner,
B/P part of the bar; another BrokenExponential component for the nuclear
disc; and a central compact, nearly circular S\'ersic component.
Table~\ref{tab:n4643-decomp} summarizes the best-fitting parameter
values, with the absolute and fractional luminosities in
Table~\ref{tab:fluxes}, and Figures~\ref{fig:n4643-bestfit-irac1} and
\ref{fig:n4643-bestfit-f160w} compare this model with the IRAC1 and
F160W data. In the lower half of Figure~\ref{fig:efits} the ellipticity
and position-angle profiles of the best-fit model are compared with
those of the data. This shows excellent agreement, even better than that
between NGC 4608 and its model, especially in terms of how the PA of the
model inside the bar agrees with the data.

The bar is, as for NGC~4608, the combination of a FlatBar component
(17\% of the total galaxy light, with a break radius of $\approx
44\arcsec$) and a Sersic\_GenEllipse component (with slightly discy
isophotes) for the B/P bulge (31\% of the total light). In contrast to
NGC~4608, the B/P bulge surface-brightness profile is closer to a
Gaussian than to an exponential (S\'ersic $n = 0.61$). As in the case of
NGC~4608, there is a slight offset between the position angles of the
FlatBar and B/P-bulge components, with the latter about 10\degr{} closer
to the galaxy major axis. Again, this agrees with the expected
projection effects acting on a bar with a vertically thin outer
component and an inner B/P component \citep{erwin-debattista13}.

Finally, the center of NGC~4643 is modeled by the combination of two
elements. The first, which we argue represents a nuclear disc, is an
elliptical BrokenExponential component with a position angle of
52.6\degr; this is beautifully consistent with the global PA of 53\degr.
(The main-disc BrokenExponential component of our model has a slightly
different PA of $\sim 59\degr$, which may represent the influence of
weak spirals in the region just outside the bar.) The ellipticity of
this component is 0.13, which is somewhat rounder than the outer disc
($\epsilon \approx 0.2$); this may indicate that the nuclear disc is
thicker than the outer disc. The profile, as noted, is a broken
exponential, with inner and outer scale lengths of $\approx 300$ and 160
pc, respectively, and a break radius of $\approx 265$ pc.

The innermost component models the steep central rise in the surface
brightness interior to $r \sim 0.5\arcsec$, using a circular S\'ersic component
with $n = 1.63$ and $\re \approx 35$ pc.

\begin{figure*}
\begin{center}
\hspace*{-1mm}\includegraphics[scale=0.505]{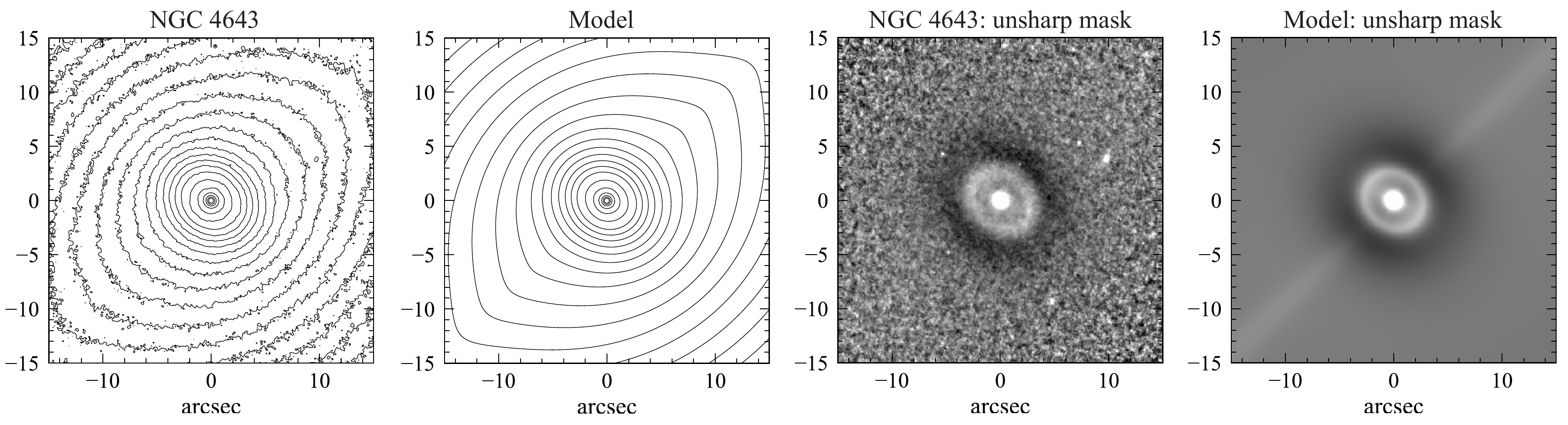}
\end{center}

\caption{Comparison of the nuclear-disc region of NGC~4643 with our
best-fitting model. \textbf{Left two panels:} Logarithmically spaced
isophotes for the F160W image of NGC~4643 (left) and our model (right).
\textbf{Right two panels:} Unsharp masks of the same (made using a
Gaussian with $\sigma = 15$ pixels). Note the clearly ringlike feature
in the unsharp masks, produced (in the model, at least) by the abrupt
change in slope of the nuclear disc rather than by a separate nuclear
ring.}\label{fig:n4643-umasks}

\end{figure*}

A question remains: can our model really explain the strikingly ringlike
appearance that unsharp masks of NGC~4643 images display?
Figure~\ref{fig:n4643-umasks} compares unsharp masks of the inner region
of NGC~4643 and the best-fitting model image, showing how the appearance
of a ring can indeed be created by the sharp break in the
broken-exponential profile of the nuclear disc. (See also
Appendix~\ref{app:nd} for evidence that the broken-exponential nature
of the nuclear disc's profile is robust against changes to how we model
this part of the galaxy.)

\begin{table}
\caption{NGC 4643: 2D Decomposition}
\label{tab:n4643-decomp}
\begin{tabular}{@{}llrl}
\hline
Component          & Parameter & Value & Units \\
\hline

 Sersic             & PA & $40 \pm 6$ & \degr{} \\
 (Bulge)            & $\epsilon$ & $0.020 \; [+0.005, -0.004]$ &  \\
                    & $n$ & $1.6 \pm 0.1$ &  \\
                    & $\mu_{e}$ & $14.3 \; [+0.1, -0.09]$ & SB \\
                    & $r_{e}$ & $0.37 \pm 0.02$ & \arcsec \\
                    &  & $35 \pm 2$ & pc \\
 BrokenExp          & PA & $52.97 \; [+0.05, -0.04]$ & \degr{} \\
 (Nuclear Disc)     & $\epsilon$ & $0.1269 \; [+0.0002, -0.0003]$ &  \\
                    & $\mu_{0}$ & $13.62 \; [+0.02, -0.01]$ & SB \\
                    & $h_{1}$ & $3.20 \; [+0.06, -0.05]$ & \arcsec \\
                    &  & $300 \; [+6, -5]$ & pc \\
                    & $h_{2}$ & $1.687 \; [+0.002, -0.001]$ & \arcsec \\
                    &  & $157.9 \; [+0.1, -0.09]$ & pc \\
                    & $R_{\rm brk}$ & $2.841 \pm 0.009$ & \arcsec \\
                    &  & $265.9 \pm 0.8$ & pc \\
                    & $\alpha$ & $12 \; [+2, -1]$ & $\arcsec^{-1}$ \\
 Sersic\_GenEllipse & PA & $123.12 \pm 0.02$ & \degr{} \\
 (bar: BP bulge)    & $\epsilon$ & $0.1534 \pm 0.0002$ &  \\
                    & $c_{0}$ & $-0.1396 \; [+0.0006, -0.0008]$ &  \\
                    & $n$ & $0.6147 \; [+0.0003, -0.0005]$ &  \\
                    & $\mu_{e}$ & $16.9134 \; [+0.0006, -0.0005]$ & SB \\
                    & $r_{e}$ & $13.349 \; [+0.006, -0.004]$ & \arcsec \\
                    &  & $1249.1 \; [+0.5, -0.4]$ & pc \\
 FlatBar            & PA & $133.276 \pm 0.004$ & \degr{} \\
 (bar: outer)       & $\epsilon$ & $0.91756 \pm 0.00003$ &  \\
                    & $\Delta$PA$_{\rm m}$ & $32.84 \; [+0.03, -0.02]$ &  \\
                    & $\mu_{0}$ & $17.5342 \; [+0.0009, -0.0007]$ & SB \\
                    & $h_{1}$ & $90.0 \pm 0$ & \arcsec \\
                    &  & $8421.2 \pm 0$ & pc \\
                    & $h_{2}$ & $5.01 \pm 0.01$ & \arcsec \\
                    &  & $469 \pm 1$ & pc \\
                    & $R_{\rm brk}$ & $44.55 \pm 0.01$ & \arcsec \\
                    &  & $4168 \pm 1$ & pc \\
                    & $\alpha$ & $0.401 \pm 0.002$ & $\arcsec^{-1}$ \\
 BrokenExp          & PA & $58.7  \pm 0.1$ & \degr{} \\
 (main disc)        & $\epsilon$ & $0.178 \pm 0.008$ &  \\
                    & $\mu_{0}$ & $19.1226 \; [+0.0008, -0.0007]$ & SB \\
                    & $h_{1}$ & $81.3  \; [+0.7, -0.6]$ & $\prime\prime$ \\
                    &  & $7600 \pm 60$ & pc \\
                    & $h_{2}$ & $10.70 \; [+0.05, -0.04]$ & $\prime\prime$ \\
                    &  & $1001 \; [+5, -4]$ & pc \\
                    & $R_{\rm brk}$ & $105.7 \pm 0.2$ & $\prime\prime$ \\
                    &  & $9890 \pm 20$ & pc \\
                    & $\alpha$ & $0.13^{1}$ & $\arcsec^{-1}$ \\


\hline 
\end{tabular}

\medskip

Summary of the final 2D decomposition of NGC~4643. Column 1: \Imfit{}
component names. Column 2: Parameter names. Column 3: Best-fit parameter
value; note that for size parameters, we include sizes in arcsec and
also in pc. Errors are nominal 68\% confidence intervals from bootstrap
resampling, and should be considered underestimates; errors for linear
sizes do \textit{not} include uncertainties in the distance. Column 4:
Units of the parameter (``SB'' = $H$-band mag arcsec$^{-2}$). For all
parameters except the final BrokenExponential (``main disc''), values
come from fitting the \textit{HST} WFC3-IR F160W image; for the latter,
values come from fitting the \textit{Spitzer} IRAC1 image.
Surface-brightness values are for $\mu_{H}$. Notes: 1 = at limit of
parameter-value boundaries; uncertainty undefined.

\end{table}

\begin{table}
\caption{Model Component Fluxes and Fractions}
\label{tab:fluxes}
\begin{tabular}{@{}llr}
\hline
Component          & $M_{H}$ & Fraction   \\
\hline
\multicolumn{3}{c}{NGC~4608}\\
\hline
Gaussian           &  $-14.74$   & 0.00056 \\
(NSC)              &   & \\
Sersic             &  $-20.57$   & 0.121   \\
(bulge)            &   & \\
Sersic\_GenEll     &  $-21.50$   & 0.286   \\
(bar: B/P bulge)   &   & \\
FlatBar            &  $-20.74$   & 0.142   \\
(bar: outer)       &   & \\
GaussianRingAz     &  $-19.17$   & 0.0333  \\
(inner ring)       &   & \\
BrokenExp          &  $-21.91$   & 0.417   \\
(main disc)        &   & \\
\hline 
\multicolumn{3}{c}{NGC~4643}\\
\hline
Sersic             &  $-17.94$   & 0.0050   \\
(bulge/NSC)        &   & \\
BrokenExp          &  $-21.51$   & 0.134   \\
(nuclear disc)     &   & \\
Sersic\_GenEll      &  $-22.42$   & 0.310   \\
(bar: B/P bulge)   &   & \\
FlatBar            &  $-21.74$   & 0.166   \\
(bar: outer)       &   & \\
BrokenExp          &  $-22.66$   & 0.386   \\
(main disc)        &   & \\
\end{tabular}

\medskip

Absolute and relative luminosities of components in our best-fitting 2D
decompositions for the two galaxies studied in this paper.  Column 1:
\Imfit{} component names (corresponding galaxy component in second row;
NSC = nuclear star cluster). Column 2: $H$-band absolute magnitude of
component (assuming $D = 17.3$ Mpc for NGC~4608 and 19.3 Mpc for
NGC~4643). Column 3: Fraction of total galaxy luminosity.

\end{table}

\subsection{Contrasting NGC~4608 and NGC~4643: Isolating the Central Bulge and Nuclear-Disc Components}

Our modeling suggests that NGC~4608 and NGC~4643 have very similar
structures -- except inside their B/P bulges. The former appears to have
something rather like a modest-sized, nearly spherical classical bulge
(ellipticity $\approx 0.04$, S\'ersic $n = 2.2$, $\re \sim 310$ pc, with
a nuclear star cluster in its center), while the latter has (in
projection) an elliptical structure with a broken-exponential profile
and a compact, spherical structure (S\'ersic $n = 1.6$, $\re \sim 35$
pc) in the very center.

\begin{figure}
\begin{center}
\hspace*{-3mm}\includegraphics[scale=0.495]{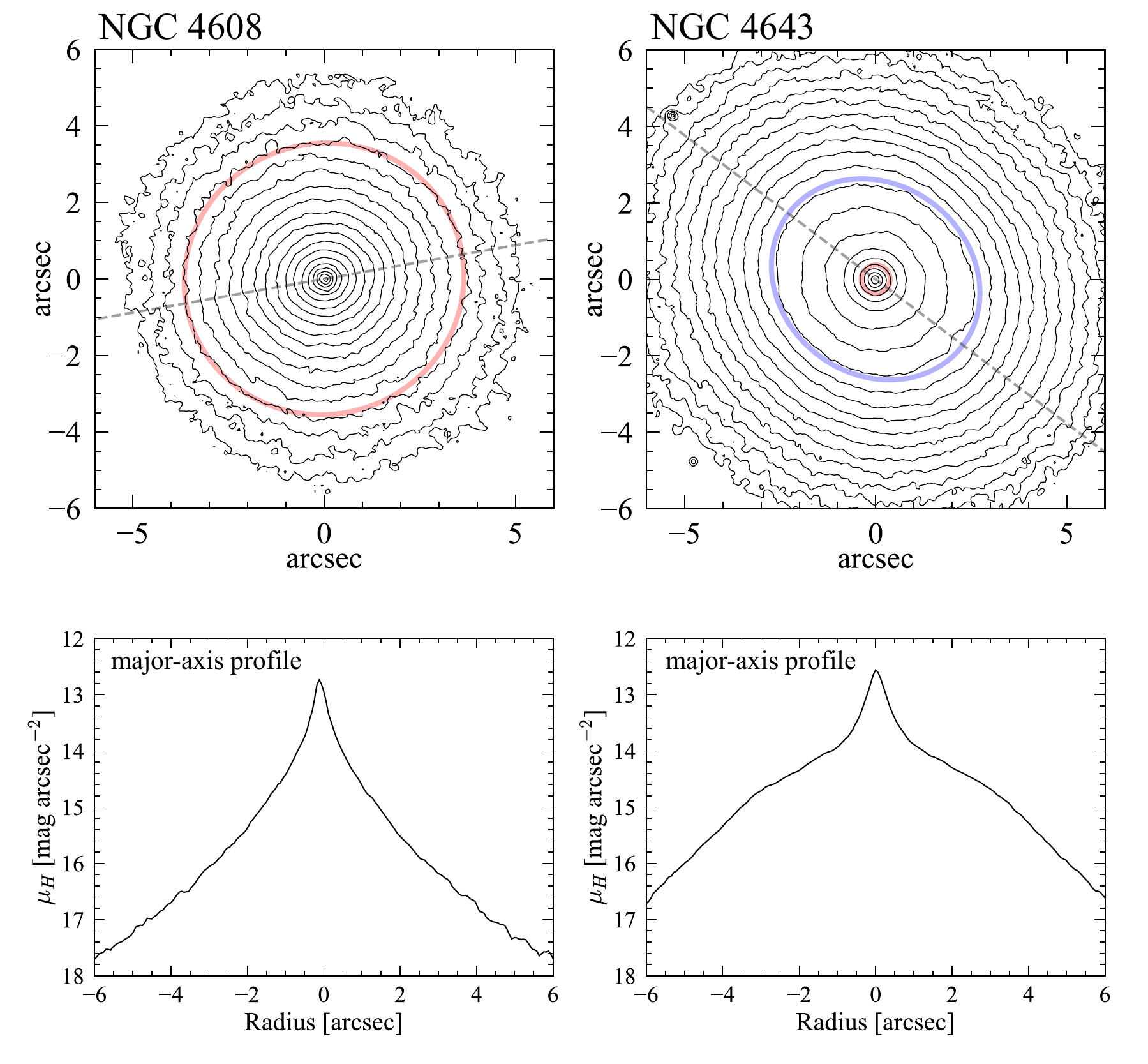}
\end{center}

\caption{Isolation of the central stellar components of NGC 4608 and NGC
4643. \textbf{Upper left:} The classical bulge (+ NSC) in NGC~4608.
Contours show logarithmically spaced isophotes from the
\textit{HST} WFC3-IR F160W image after subtracting the best-fitting
model components corresponding to the disc and bar. The red circle marks
the effective radius ($\re = 3.67\arcsec = 310$ pc) of the S{\'e}rsic
component from the full fit; the dashed grey line indicates the galaxy
major axis. \textbf{Upper right:} Same, except now showing the nuclear
disc and compact classical bulge/NSC in NGC~4643. The blue ellipse marks the
break radius of the broken-exponential component from the full fit
($a_{\rm brk} = 2.86\arcsec \approx 270$ pc), while the small red circle marks
the effective radius ($\re = 0.38\arcsec \approx 35$ pc) of the inner
S{\'e}rsic component from the same fit. \textbf{Bottom panels:} Profiles
along the major axis through both isolated components.}
\label{fig:central-components}

\end{figure}

These parameters are taken from our best-fitting models; to see what the
centers might look like if we could remove the rest of the galaxy
surrounding them, Figure~\ref{fig:central-components} shows isophotes
and major-axis profiles from the F160W image after subtracting the disc
and bar components of our best-fitting models.

\section{The Stellar Kinematics of the Central Structures}\label{sec:kinematics}

The preceding decompositions indicate that both NGC~4608 and NGC~4643
have central regions ($r \la 1.5$ kpc) dominated by the B/P bulges of
their bars. In the inner few hundred parsecs, however, the galaxies
differ significantly. Our preceding, photometry-based claim -- that
NGC~4608 hosts a modest classical bulge while NGC~4643 has instead a
nuclear disc -- implies rather different stellar kinematics: we would
expect a ``classical bulge'' to be slowly rotating and dominated by
velocity dispersion, while a ``nuclear disc'' ought to be rapidly
rotating, with lower velocity dispersion. In addition, our argument that
the mildly elongated isophotes at $r \sim 1$ kpc in both galaxies are
due to the B/P bulges of bars -- rather than to massive, large-scale
classical bulges -- suggests that the kinematics there should be barlike
rather than spheroidal. In this section, we use archival and published
IFU data to probe the stellar kinematics in the inner regions of both
galaxies to test these predictions.

\subsection{SAURON Stellar Kinematics}

We begin with published stellar kinematics from observations with the
SAURON integral field unit. Although VLT-MUSE data with superior
coverage and spatial and spectral resolution is available for NGC~4643,
the SAURON data is all we have for NGC~4608, so it makes sense to
compare the kinematics from the SAURON observations of the two galaxies
first. We can then see how well these compare with the MUSE data for
NGC~4643.

\subsubsection{NGC 4608} 

\begin{figure*}
\begin{center}
\hspace*{-3mm}\includegraphics[scale=0.73]{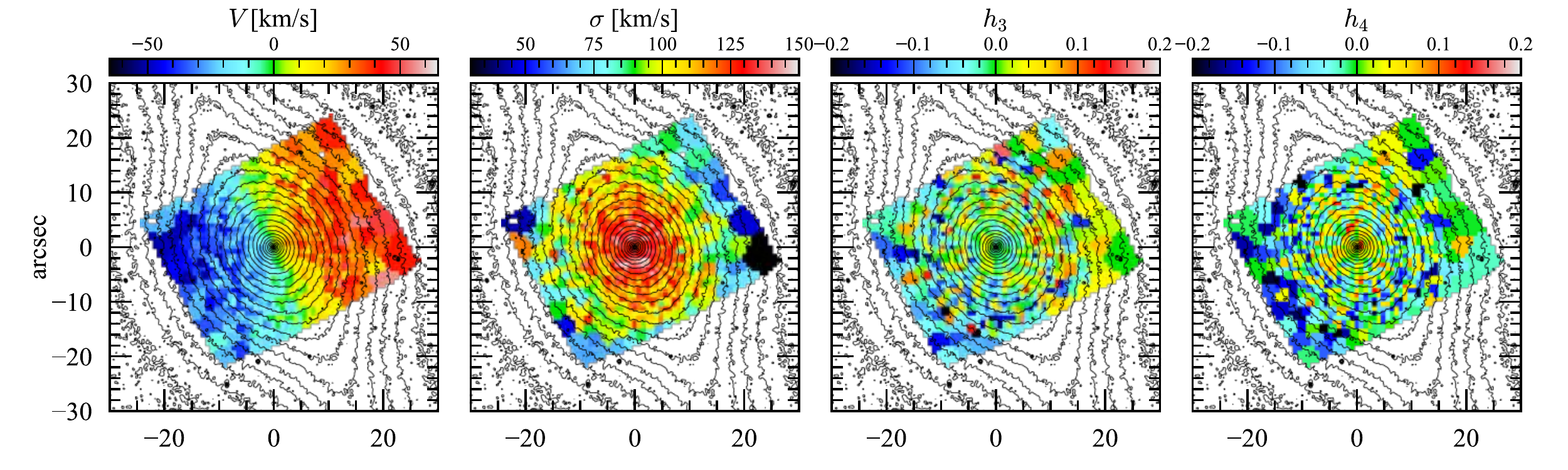}
\end{center}

\caption{NGC 4608 SAURON kinematic maps, with data from \atlas{} (colour)
plotted on top of \textit{HST} F160W isophotes. In the \hthree{} panel,
there is evidence for a weak asymmetry: negative \hthree{} on the left
side, positive \hthree{} on the right. This matches the velocity pattern
in the first panel; such a $V$-\hthree{} correlation is a signature of
bar orbits.}\label{fig:n4608-kinematics}

\end{figure*}

Figure~\ref{fig:n4608-kinematics} shows the SAURON stellar kinematics
for NGC~4608: stellar velocity, velocity dispersion, \hthree, and
\hfour, along with isophotes from our F160W image to show the underlying
stellar structure. We can see a clear pattern of rotation, along with
velocity dispersion that increases toward the center of the galaxy; the
dispersion appears elongated along the major axis of the bar, though it
becomes rounder in the very center. The \hthree{} map shows relatively
little structure, though there is weak evidence for a slight
\textit{correlation} between velocity and \hthree{}, with positive
velocities and \hthree{} values on the NW side of the galaxy center and
negative values of both on the SE side. This is potentially significant,
since $V$-\hthree{} correlation is a prediction of bar models
\citep[e.g.,][]{iannuzzi15,li18} due to the presence of elliptical,
bar-supporting orbits; this suggests that the region from $r \sim
5$--20\arcsec{} along the bar minor axis is dominated by bar orbits.

\begin{figure*}
\begin{center}
\hspace*{-3mm}\includegraphics[scale=0.73]{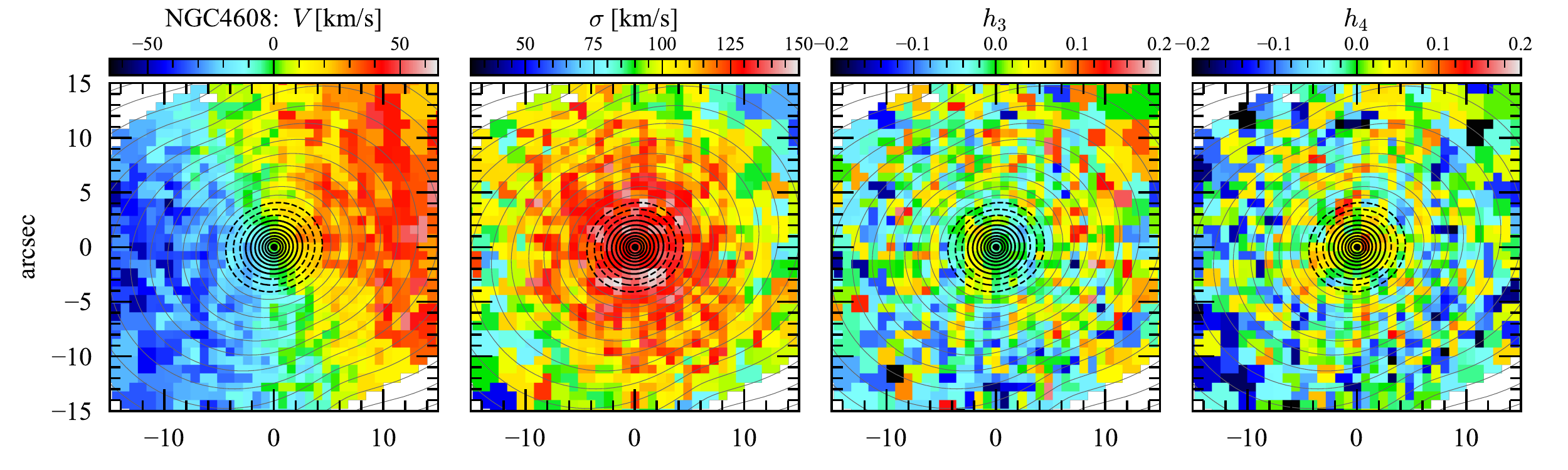}
\end{center}

\caption{A close-up of Figure~\ref{fig:n4608-kinematics}, but with
contours now showing the ratio of the classical-bulge (S\'ersic)
component to the rest of our best-fitting model of NGC~4608 (e.g.,
Figure~\ref{fig:n4608-bestfit-f160w}). Solid black contours indicate
where the classical-bulge component is brighter, dashed black contours
where its brightness is 50--100\% of the rest of the model, and thin
grey contours where it is fainter. The region dominated by the
bulge component has almost no rotation, centrally increasing velocity
dispersion, and little or no signature in \hthree{} or \hfour; this
is suggestive of a classical bulge.}\label{fig:n4608-kinematics-ratio}

\end{figure*}

In Figure~\ref{fig:n4608-kinematics-ratio} we show a close-up of the same
kinematic data. Here, instead of the observed F160W isophotes, we plot
isophotes which show the \textit{ratio} between the classical-bulge
component in our best-fitting \Imfit{} model and the rest of the model.
The solid lines show where the former component is brighter than the
rest of the model; this is where the classical-bulge component is the
dominant stellar structure in our model. This region ($r \la
2.5\arcsec$) shows weak rotation, uniformly high velocity dispersion,
and a tenuous hint of weak $V$-\hthree{} \textit{anticorrelation} (though
the statistical significance of the latter is dubious). All of this is
consistent with a slowly rotating classical bulge.

\subsubsection{NGC 4643} 

\begin{figure*}
\begin{center}
\hspace*{-8mm}\includegraphics[scale=0.98]{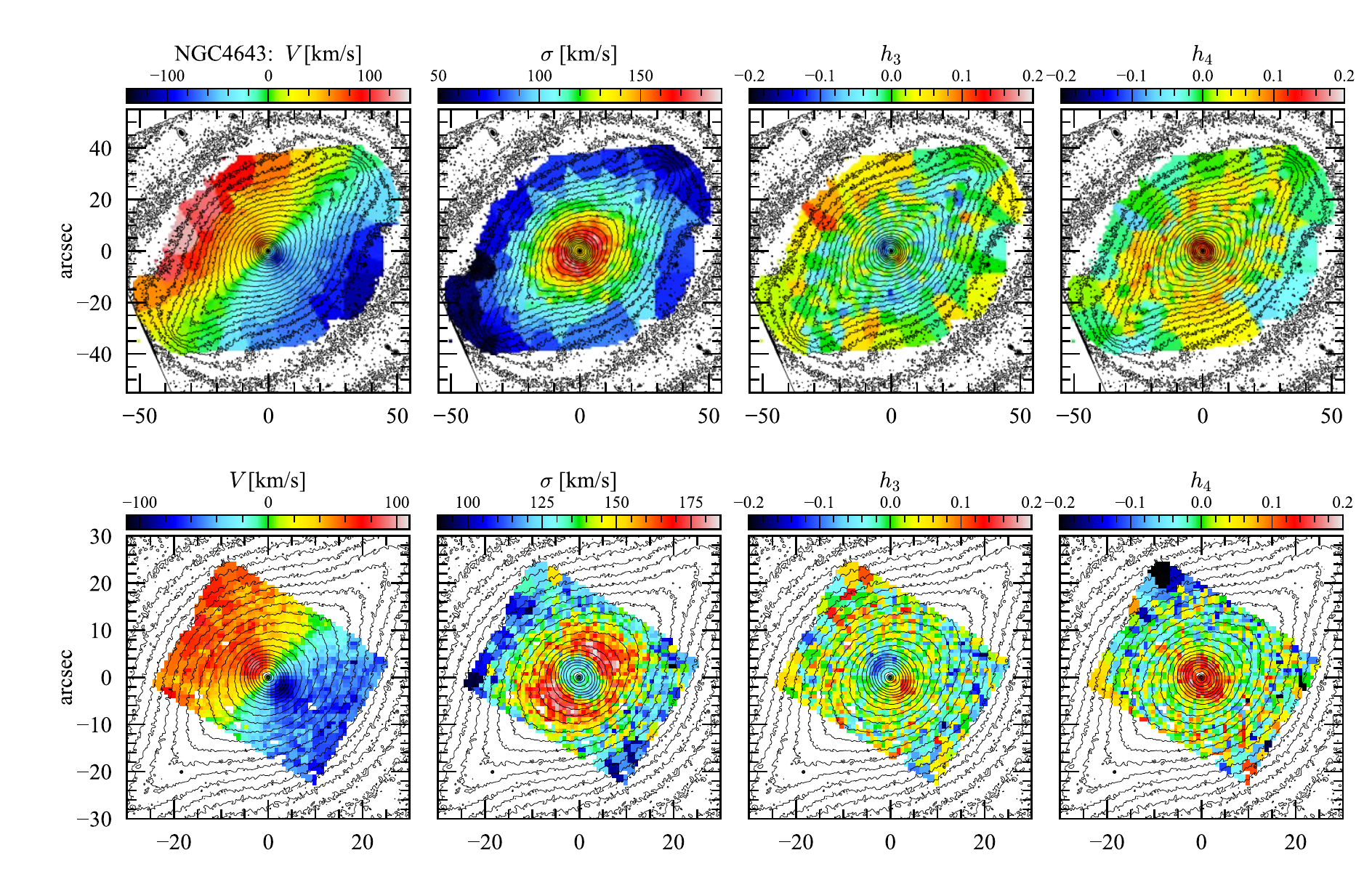}
\end{center}

\caption{NGC 4643 SAURON kinematic maps, plotted on top of \textit{HST} F160W
isophotes. Top panels show maps from \citet{seidel15b},
bottom panels show maps from \atlas{}. Note the rapid rotation in the inner
$r \la 5\arcsec$, accompanied by \textit{lower} velocity dispersion, $V$-\hthree{}
anticorrelation, and positive \hfour. (A very weak indication of $V$-\hthree{}
\textit{correlation} can be seen at $r \sim \pm 20\arcsec$ along the minor axis
of the bar, suggesting the presence of bar orbits.)}\label{fig:n4643-kinematics}

\end{figure*}

Here, we examine the SAURON velocity fields for NGC~4643.
Figure~\ref{fig:n4643-kinematics} shows SAURON data for this galaxy,
along with isophotes from our F160W image of the galaxy. In addition to
the \atlas{} kinematics (lower panel), we also show (top panel)
kinematics from the study of \citet{seidel15b}, which used multiple
pointings to sample a larger field of view.

Outside $r \sim 6\arcsec$, the stellar-velocity pattern is similar to
that in NGC~4608: clear stellar rotation, an elongated region of higher
velocity dispersion aligned with the bar and increasing toward the
center, and a tenuous $V$-\hthree{} correlation (extending to $r \sim
25\arcsec$ along the major axis of the galaxy, which is approximately
the minor axis of the bar). As with NGC~4608, the evidence for
$V$-\hthree{} correlation in the bar is very weak; however, for this
galaxy the VLT-MUSE kinematics provide clear confirmation of this
pattern (e.g., \citealt{gadotti19} and Section~\ref{sec:n4643-muse}).

But for $r \la 6\arcsec$ ($r \la 500$~pc), the stellar kinematics is
very different from that of NGC~4608. There is a region of rapid
rotation accompanied by a clear \textit{anticorrelation} with \hthree.
There is also a clearly elliptical region of distinctly \textit{lower}
velocity dispersion, and a similar region of elevated \hfour.

\begin{figure*}
\begin{center}
\hspace*{-3mm}\includegraphics[scale=0.73]{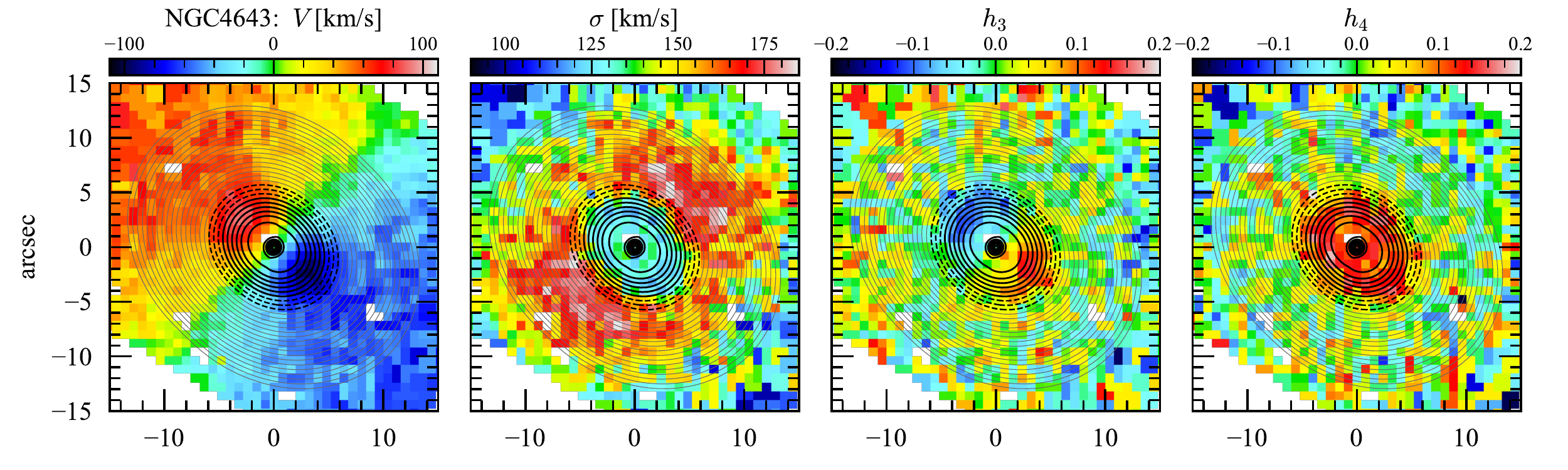}
\end{center}

\caption{A close-up of the lower panels (\atlas{} stellar kinematics) of
Figure~\ref{fig:n4643-kinematics}, but with contours now showing the
ratio of the nuclear-disc component to the rest of our best-fitting
model of NGC~4643 (e.g., Figure~\ref{fig:n4643-bestfit-f160w}). Solid
black contours indicate where the nuclear disc is brighter, dashed black
contours where its brightness is 50--100\% of the rest of the model, and
thin grey contours where it is fainter. Note that the region dominated
by the nuclear-disc component matches almost perfectly with the fast
rotation, \textit{lower} velocity dispersion, strong $V$-$\hthree$
anticorrelation, and enhanced \hfour.}\label{fig:n4643-kinematics-ratio}

\end{figure*}

From the bottom panel of Figure~\ref{fig:n4643-kinematics}, it is clear
that this inner region of rapid, disclike rotation is associated with
the elliptical inner isophotes we previously identified as due to a
nuclear disc (e.g., Section~\ref{sec:n4643-fit}).
Figure~\ref{fig:n4643-kinematics-ratio} demonstrates this more clearly
by comparing the \atlas{} kinematics with the isophotes from an image
which is the ratio of the nuclear-disc component of our model to the
rest of the model (similar in spirit to
Figure~\ref{fig:n4608-kinematics-ratio}). Solid contour lines show where
the nuclear-disc component is the dominant component, and this is
precisely where we see the rapid rotation, $V$-\hthree{}
anticorrelation, lower dispersion, and elevated \hfour.

Our conclusion is that the inner $r \la 500$~pc of NGC~4643 is dominated
by the rapid rotation, $V$-\hthree{} anticorrelation, and lower
dispersion typical of a disc rather than a classical bulge. The only
really puzzling part of this is the rather high \hfour{} in the
nuclear-disc region, since a pure disc should probably have \hfour{}
close to zero \citep[see, e.g.,][]{debattista05}. As we will argue in
Section~\ref{sec:n4643-muse}, this is probably due to the superposition
of two components along our line of sight: the dominant, low-$\sigma$
nuclear-disc kinematics and the innermost part of the bar's B/P bulge,
with higher $\sigma$.

\subsection{NGC 4643: Additional Evidence from MUSE}\label{sec:n4643-muse}

NGC~4643 is one of the barred galaxies studied by the TIMER project
\citep{gadotti19}. Fig.~3 of that paper shows the VLT-MUSE-based stellar
kinematics for this galaxy, and evinces the same stellar-kinematic
patterns we noted in the SAURON data, albeit with higher spatial
resolution and considerably better $S/N$. In particular, the
nuclear-disc region is singled out by a pattern of high rotation, clear
$V$-\hthree{} anticorrelation, lower velocity dispersion, and positive
\hfour. Outside the nuclear disc, the velocity dispersion is clearly
higher along the major axis of the bar. In addition, the maps show that
near the minor axis in the outer part of the bar (e.g., between the
fourth or fifth and the seventh or eighth contour lines in their figure)
\hthree{} is \textit{correlated} with the velocity, as hinted at in the
SAURON data. The combination of the latter two features is, as
\citet{gadotti19} noted, strong kinematic evidence for a B/P bulge
within a bar.

\begin{figure*}
\begin{center}
\hspace*{-3mm}\includegraphics[scale=0.9]{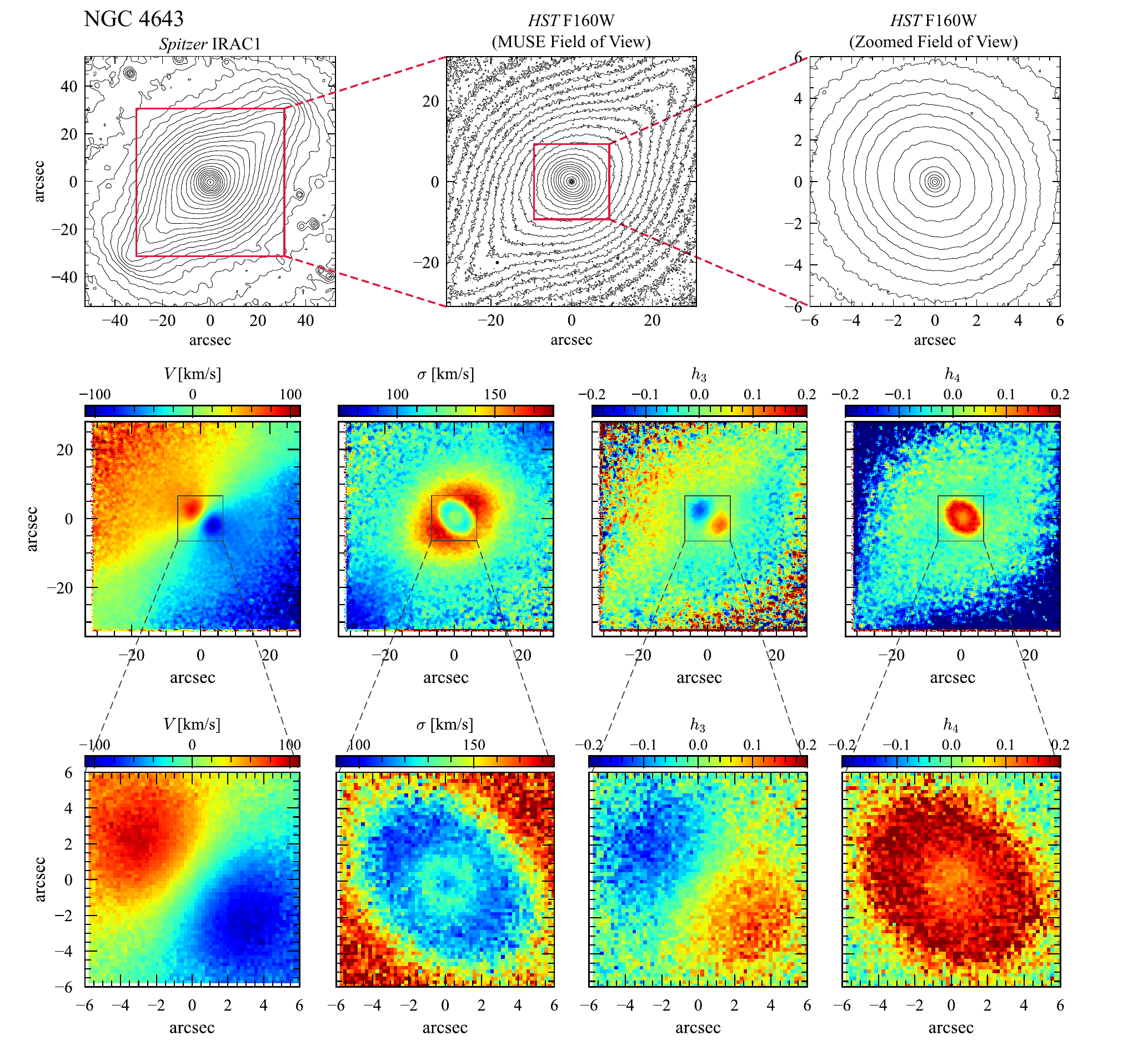}
\end{center}

\caption{MUSE stellar kinematics for NGC~4643.
\textbf{Top panels}: Isophotes from IRAC1 and F160W images of NGC~4643.
\textbf{Middle panels}: Median-smoothed stellar kinematics, from
single-spaxel analysis of VLT-MUSE data (full MUSE field of view). \textbf{Bottom panels:} Close-up
of unsmoothed stellar kinematics in the nuclear-disc region.}\label{fig:n4643-kinematics-muse-large}

\end{figure*}

In Figure~\ref{fig:n4643-kinematics-muse-large}, we show unbinned
(single-spaxel) stellar kinematics derived from the same MUSE data,
showing both the full MUSE field of view and the region containing the
nuclear disc. Here, we can see that the nuclear-disc region has two
extra kinematic features. The first is the fact that at smaller radii
(e.g., $r \la 2\arcsec$), the velocity dispersion increases to $\sigma
\sim 125$--130 \kms, versus $\sigma \sim 110$--115 kms{} in the region
$r \sim 1.5$ to 3.5\arcsec. This corresponds with a nearly circular
region of slightly lower \hfour{}.

The second interesting feature is that the dispersion drops again in the
very center ($r < 0.5\arcsec \approx 45$ pc), to a value of $\sim
110$--115 \kms. This corresponds almost exactly with the round inner
S\'ersic component in our decomposition. We note that there is some
evidence for nuclear stellar-velocity-dispersion depressions on similar
scales in other early-type spirals. In NGC~3368, which has a
similar-sized ($\re \approx 25$~pc) compact ``classical-bulge''
component \citep{nowak10,erwin15a}, the velocity dispersion observed
with VLT-SINFONI in AO mode shows a drop in the dispersion in the inner
$0.5\arcsec \approx 25$ pc (\citealt{nowak10}, their Fig.~14).
\citet{erwin15a} argued for a compact central stellar component with $\re
\approx 35$~pc in NGC~1068, where the SINFONI AO stellar kinematics of
\citet{davies07} showed a central decrease (albeit to an implausible
value of 0 \kms{} at the center, presumably due to problems subtracting
the contribution of the Seyfert nucleus). Finally, although we do not
(currently) have any evidence for a similar round, compact structure in
NGC~1097, the SINFONI AO data of \citet{davies07} shows a clear drop
from from 150 \kms{} to 100 \kms{} for $r < 0.5\arcsec \approx 35$ pc.

So there is some evidence for central drops in the stellar velocity dispersion
in the inner 25--35 pc of massive, early-type spirals. The underlying
dynamical origin of the phenomenon needs further investigation.

\begin{figure*}
\begin{center}
\hspace*{-2.5mm}\includegraphics[scale=0.73]{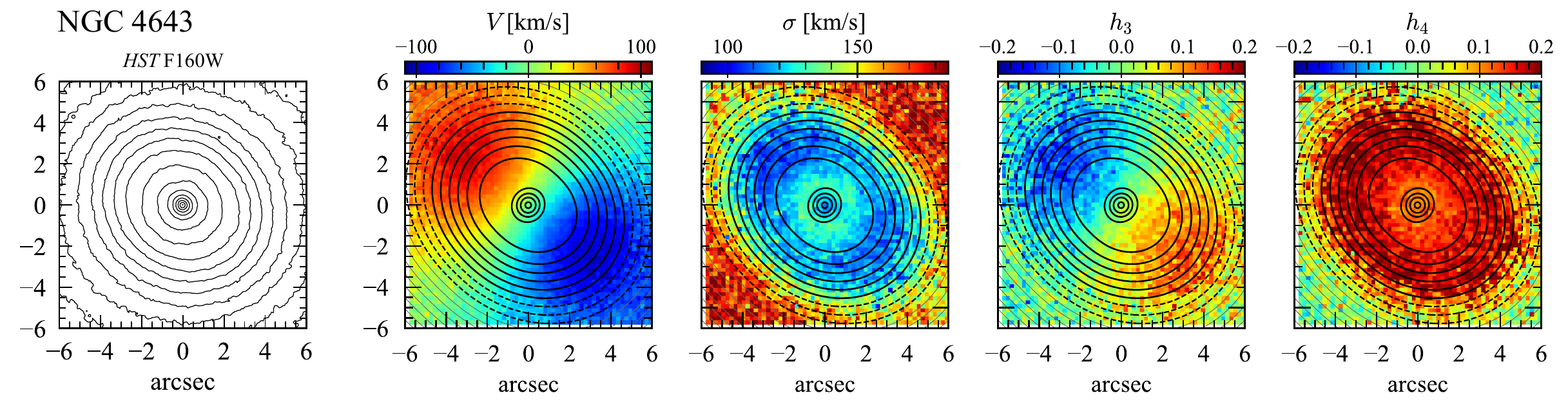}
\end{center}

\caption{MUSE stellar kinematics for the nuclear-disc region of
NGC~4643. \textbf{Leftmost panel}: Isophotes from F160W image of
NGC~4643. \textbf{Middle and right panels}: Stellar kinematics for
NGC~4643, from single-spaxel analysis of VLT-MUSE data. Contours in
these panels are as in Figure~\ref{fig:n4643-kinematics-ratio}: solid
contour lines indicate where the nuclear-disc component of our
best-fitting model is brighter, dashed black contours where its
brightness is 50--100\% of the rest of the model, and thin grey contours
where it is fainter (note that the inner four solid contour lines indicate
\textit{decreasing} levels of relative brightness toward the center, due
to the increasing importance of the compact classical-bulge/NSC component).
The nuclear-disc ratio image has been convolved to match the FWHM
$\approx 0.75\arcsec$ seeing of the MUSE kinematics.}\label{fig:n4643-kinematics-muse-ratio}

\end{figure*}

Figure~\ref{fig:n4643-kinematics-muse-ratio} shows the inner MUSE
stellar kinematics again, this time with the nuclear-disc/model ratio
isophotes, as in Figure~\ref{fig:n4643-kinematics-ratio}. From this, we
can again see that the elliptical region of high velocities, low
dispersion and anticorrelated \hthree{} matches almost perfectly with
where the nuclear-disc component dominates in our model. Fitting the
velocity field in this region using the kinemetry code of
\citet{krajnovic06} gives a kinematic position angle of $52.0\degr \pm
0.9\degr$, identical within the uncertainties with the position angle of
the BrokenExponential component we use to model the nuclear disc
(52.6\degr, Table~\ref{tab:n4643-decomp}).

\begin{figure}
\begin{center}
\hspace*{-3mm}\includegraphics[scale=0.58]{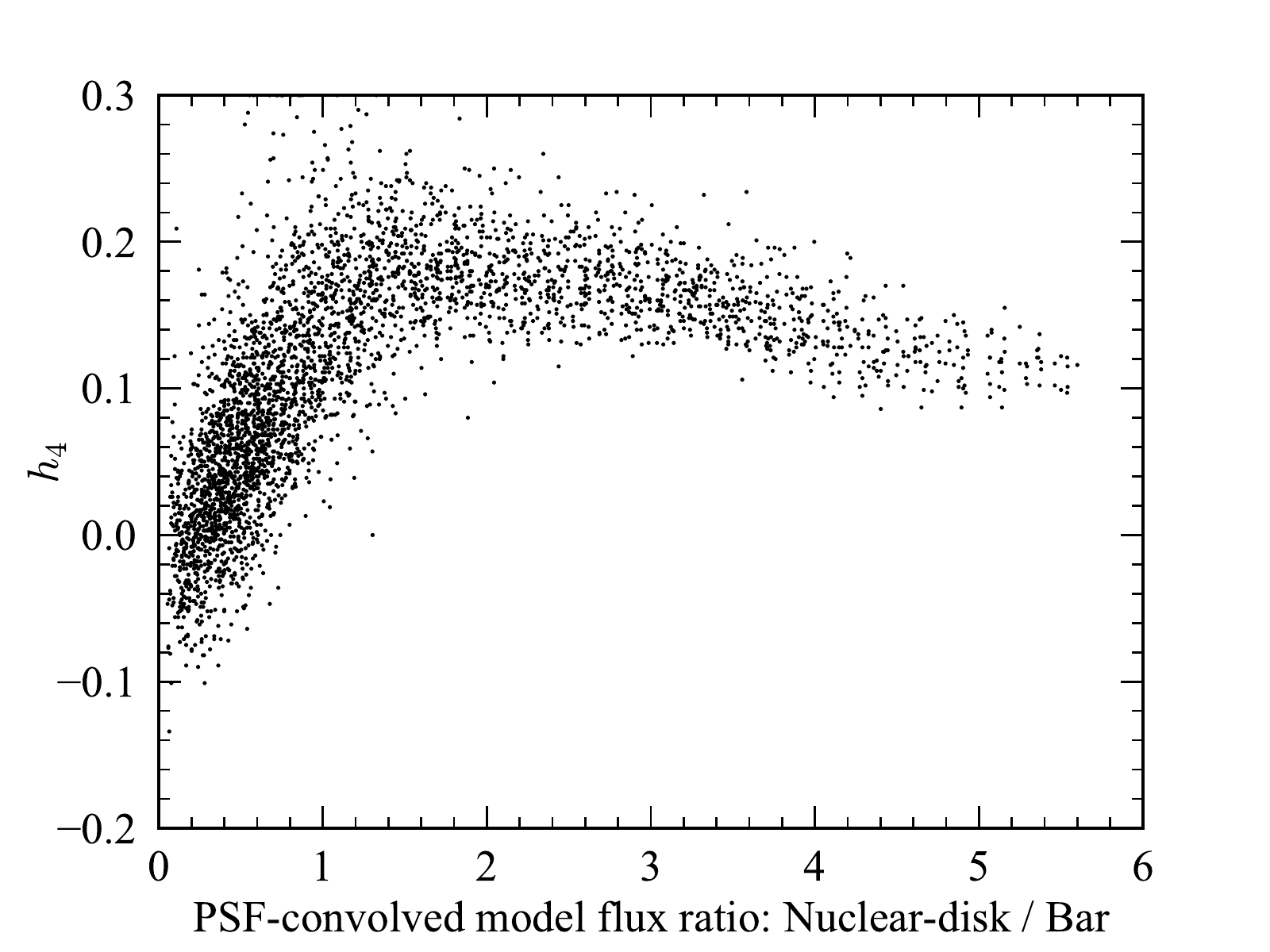}
\end{center}

\caption{MUSE stellar-kinematic \hfour{} values for $12 \times
12\arcsec$ central region of NGC~4643, plotted against the local ratio
of our nuclear-disc model to the rest of the bar model. The \hfour{}
values peak where the ratio is $\sim 1$, suggesting that the high kurtosis
is the result of nearly equal contributions from nuclear-disc and bar
LOSVDs with significantly different dispersions.}\label{fig:n4643-h4-vs-ndratio}

\end{figure}

This is also where the elevated \hfour{} is found. By plotting the
\hfour{} values against the ratio of the nuclear-disc component to the
bar model in Figure~\ref{fig:n4643-h4-vs-ndratio}, we can see a clear
trend: the kurtosis values increase as the ratio increases, peaking
where the ratio is $\sim 1.3$, and then decline as one moves to even
higher ratio values. We interpret this as the result of changing
relative contributions from the B/P bulge (high velocity dispersion) and
from the nuclear disc (low velocity dispersion) along the line of sight.
Outside the nuclear disc (ND ratio $< 1$), the LOSVD is dominated by the
high-dispersion B/P bulge LOSVD, and so the kurtosis is low. In the
region where both the B/P bulge and the nuclear disc contribute almost
equally, the kurtosis reaches a maximum. At smaller radii, where the
nuclear disc dominates, the kurtosis declines.

We will explore this phenomenon further in a follow-on paper.

\subsection{Summary of Stellar Kinematics Results}

We can describe the results of our analysis of the stellar kinematics
data from published SAURON observations (both galaxies) and from MUSE
observations (NGC~4643 only) as follows:

\begin{itemize}

\item The outer isophotes of the bars in both galaxies show rotation with
possible (NGC~4608) or clear (NGC~4643) $V$-\hthree{} correlation, as
expected for bars.

\item The B/P-bulge regions (the vertically thickened inner parts of the
bars) of both galaxies show higher velocity dispersion, aligned with the
projected B/P bulges and increasing towards the center.

\item In NGC~4608, the region we photometrically identified as a
possible classical bulge ($r \la 2.5\arcsec \approx 200$ pc) shows slow
rotation, high velocity dispersion ($\sim 140$ \kms), and no clear
$V$-\hthree{} correlation \textit{or} anticorrelation, consistent with a
slowly rotating classical spheroid.

\item By contract, in NGC~4643 the inner region that we photometrically
identified as a nuclear disc ($r \la 5\arcsec \approx 500$ pc) shows up
as an elliptical zone of significantly \textit{lower} dispersion, with
rapid rotation and strong $V$-\hthree{} anticorrelation, consistent with
a rotation-dominated disc. The dispersion has a minimum value of $\sim
110$ \kms, increasing mildly toward the center.

\item The nuclear-disc region of NGC~4643 also shows very high \hfour{}
values, declining toward the center; we tentatively interpret this as a
signature of overlapping LOSVDs from the B/P bulge (higher dispersion)
and the nuclear disc (lower dispersion), with the decline in \hfour{}
toward the center happening as the nuclear-disc component becomes more
dominant (and possibly also as the dispersion of the nuclear-disc
component becomes higher).

\item Finally, there is a drop to lower dispersion in the very innermost
region of NGC~4643 ($r < 0.5\arcsec \approx 45$ pc, from $\sim 130$
\kms{} at the edge of this region to $\sim 110$ \kms{} in the center),
coincident with the photometrically identified compact classical bulge
(or large NSC). Although there are some similar near-nuclear dispersion
drops seen in a few other massive spirals, the cause of this remains
unclear.

\end{itemize}

\section{Discussion}\label{sec:discussion}

Outside of the central few hundred parsecs, the two galaxies studied in
this paper are remarkably similar. In both cases, their bars amount to
$\sim 45$\% of the stellar light, with dominant B/P bulges comprising
$\sim 30$\% of the light. They both have distinct stellar structures
inside the B/P bulges, each of which is $\sim 12$--13\% of the total light.
But while this structure in NGC~4608 is nearly circular, centrally
concentrated (S\'ersic index $\sim 2.2$), slowly rotating, and dominated
by velocity dispersion -- matching a traditional classical bulge -- the
inner structure of NGC~4643 is a nuclear disc: elliptical and aligned
with the outer disc, with rapid rotation and lower velocity dispersion.

The photometric and kinematic evidence for a nuclear disc inside the B/P
bulge of NGC~4643 is a clear demonstration of a pattern
\citet{chung-bureau04} and \citet{bureau06} suggested based on their
studies of edge-on disc galaxies: that the B/P bulges of barred galaxies
are frequently accompanied by nuclear discs. The case of NGC~4608, on the
other hand, demonstrates that this is not a \textit{universal}
pattern: some B/P bulges contain structures much more like classical
bulges, rather than nuclear discs.

\subsection{The Nuclear Disc in NGC 4643}

The prominent nuclear disc in NGC~4643 makes up 13\% of the galaxy's
near-IR light. Assuming no strong gradients in $M/L$ and a total galaxy
stellar mass of $6.2 \times 10^{10} \Msun$ (Section~\ref{sec:overview}),
the nuclear disc has a stellar mass of $\sim 7 \times 10^{9} \Msun$.
This is about seven times the mass estimated for Milky Way's nuclear
disc \citep[$\sim 1 \times 10^{9} \Msun$;][]{nogueras-lara19,sormani20}.
The stellar-population analysis of the MUSE data in
\citet{bittner20} indicates that the nuclear disc is \textit{old} ($\sim
10.5$ Gyr) -- but slightly less so than the stars in the B/P bulge
($\sim 11.5$ Gyr) in which it is embedded. It is also slightly more
metal-rich than the B/P bulge \citep{seidel15b,gadotti19,bittner20}.
This suggests that it was formed not long after the bar itself, from gas
that was slightly enriched compared to that from which the stars in the
bar were formed.

The most peculiar thing about NGC~4643's nuclear disc is its
surface-brightness profile: instead of the standard single-exponential
form one (perhaps naively) expects of a disc, it has a downbending,
broken-exponential profile (a Freeman Type~II profile, in the scheme
used for large-scale discs) -- with a break which is sufficiently sharp
to appear ringlike in unsharp masks (e.g.,
Figure~\ref{fig:n4643-umasks}); see Appendix~\ref{app:nd}. We are
not aware of any similar structures in other galaxies, though it is
quite possible they have escaped notice because few nuclear discs have
been analyzed with sufficient resolution or in sufficient detail.
Broken-exponential profiles are, of course, quite common in large-scale
discs, but it is unclear whether the proposed mechanisms for those discs
-- e.g., radial truncations in star-formation in concert with
spiral-driven radial migration \citep[e.g.,][and references
therein]{debattista17c} -- would apply here. One speculative possibility
is that this is an end-state for the combination of a nuclear disc and a
nuclear ring, with an initially narrow ring having broadened over time,
blending with the nuclear disc to produce the broken-exponential
structure.

Most theoretical discussions of nuclear rings and discs inside bars
suggest that the stars on them form from gas which has settled onto the
x2 orbits of a bar potential, which implies that the stars should
themselves be trapped around the same family of orbits. Since x2 orbits
are elongated perpendicular to the x1 orbits that support the bar, we
might expect nuclear discs to themselves be elongated perpendicular to
their host bars. For the case of NGC~4643, where the bar has a PA of
133\degr, the line of nodes is at 53\degr, and the inclination is
38\degr, the minor axis of the bar should have an observed PA of
47\degr; an elliptical structure oriented along this axis would have an
observed PA in between that and line of nodes. However, the ellipse fits
to the observed isophotes in the nuclear-disc region
(Figure~\ref{fig:efits} and the nuclear-disc component in our
best-fitting model of the F160W image (Table~\ref{tab:n4643-decomp})
both agree on a PA of 52--53\degr, essentially identical to that of the
main disc -- and to the kinematic PA of the nuclear-disc region
(52\degr; Section~\ref{sec:n4643-muse}). The implication is that the
nuclear disc is intrinsically close to circular in the plane of the
galaxy, rather than significantly elongated perpendicular to the bar.

The observed ellipticity of the nuclear disc is $\approx 0.13$, which is
rounder than that of the main disc (0.20). This in turn suggests that
the nuclear disc, if circular, is intrinsically somewhat thicker than
the main disc -- or perhaps that it is slightly elongated
\textit{parallel to} the bar. The latter possibility might mean that
the stars are trapped around round inner x1 orbits of the bar, and
even that the nuclear disc could predate the bar. As noted above, the stellar
ages estimated from the analysis of the MUSE data by \citealt{gadotti19}
show that the stars in the nuclear disc are approximately the same age as 
the stars in the bar surrounding it (mass-weighted ages $\sim 10$ Gyr);
given the inherent uncertainties in age estimates for older stellar
populations, we probably cannot tell whether the nuclear disc is truly
younger or older than the bar.

\subsection{Speculations About Formation Histories}

We have emphasized that both NGC~4608 and NGC~4643 are quite similar in
their large-scale morphology: massive, early-type discs with very strong
bars that contain prominent B/P bulges. But their inner structure is
rather different: NGC~4608 has what appears to be a modest,
kinematically hot classical bulge (and a very compact NSC), while
NGC~4643 has a kinematically cool nuclear disc and what is either a very
compact classical bulge or a large NSC. This suggests two questions.
First, why does NGC~4608 have a significant classical bulge, while
NGC~4643 does not? Second, why is there a nuclear disc present in
NGC~4643 but not in NGC~4608?

The formation of a classical bulge in NGC~4608 could potentially be
ascribed to local variations in the very early formation stages, leading
to more merger activity in the central region of one galaxy than the
other. The question then becomes why NGC~4608 was not able to form
a nuclear disc, despite having a strong bar very much like NGC~4643's.

One possibility is that this dichotomy reflects the different
environments of the two galaxies: NGC~4643 is a field galaxy, while
NGC~4608 is a member of the Virgo Cluster. In this scenario, we could
argue that NGC~4608 could have undergone ram-pressure stripping within
the cluster environment early enough in its history that its bar was
unable to drive significant gas inflow and form a nuclear disc.
NGC~4643, by contrast, must have retained enough gas after its bar
formed for bar-driven inflow and star formation to have formed its
nuclear disc. NGC~4643 in fact still retains some gas, with \hi{}
detections \citep[e.g.,][]{van-driel00} and H$\alpha$ emission within
the bar which is consistent with the optical dust lanes
(Figure~\ref{fig:n4643-overview}) and with kinematics consistent with
that of the stars \citep{gadotti19}.

We should note, however, that another barred S0 in the Virgo Cluster --
NGC~4371 -- contains evidence for both a modest classical bulge
\citep{erwin15a} \textit{and} a significant nuclear disc/ring with a
slightly younger and more metal-rich population in its MUSE spectra
\citep{gadotti15,bittner20}. So it is clear that the cluster environment
doesn't exclude nuclear-disc formation.

\subsection{Classical Bulges and B/T Ratios}\label{sec:bt-ratios} 

As mentioned in the Introduction, large samples of galaxies, imaged at
moderate to low resolution, are typically analyzed in terms of simple
bulge/disc decompositions, treating the galaxies as consisting of just
an (exponential) disc and an optional (S\'ersic-profile) bulge; the
latter is often assumed to be a classical bulge, or else treated in a
dichotomous fashion as either a classical bulge \textit{or} a
pseudobulge (e.g., depending on whether the S\'ersic index $n$ is greater
or less than 2). Given the complexity in the central regions we find for
these two galaxies, it is instructive to consider how they would be
analyzed as part of a large, low-spatial-resolution sample.

To do this, we performed simple bulge/disc decompositions with \Imfit{}
on two sets of images. The first was the same \textit{Spitzer} IRAC1
images (Section~\ref{sec:other-imaging-data}) used as part of our
multi-component decompositions in Section~\ref{sec:decomp}. For
NGC~4643, this can be compared with the two-component fit done to the
same image by \citet{salo15}. The second sets of images consisted of
SDSS $r$-band images, artificially redshifted to $z = 0.04$ (chosen as a
redshift typical of large, SDSS-based samples); this corresponds to an
angular-diameter distance almost exactly ten times further away than the
galaxies' actual distances. We redshifted each image by first convolving
it with a Gaussian so that it had same effective resolution it would
have under the original observing conditions, but with the galaxy at $z =
0.04$; we then binned the pixels 10:1. We did not attempt to include any
$k$-correction or surface-brightness dimming, since these are minor
effects for the target redshift.

For NGC~4608, the two-component fit of the IRAC1 image gives $B/T =
0.49$, with S\'ersic $n = 2.4$ and $\re \approx 1.1$ kpc for the bulge
component. Decomposition of the redshifted $r$-band image also gives
$B/T = 0.49$; the S\'ersic component now has a higher value of $n = 3.8$
but essentially the same \re.

For NGC~4643, we find even more ``bulge-dominated'' fits, with $B/T =
0.65$ (S\'ersic $n = 2.4$, $\re \approx 1.3$ kpc) for the IRAC1 image;
this is similar to the two-component fit to the same image by
\citet{salo15}: $B/T = 0.72$, S\'ersic $n = 3.0$ and $\re \approx 2.3$
kpc.\footnote{Retrieved from webpage
\url{https://www.oulu.fi/astronomy/S4G_PIPELINE4/P4STORE/}} For the
redshifted $r$-band image, we find $B/T = 0.61$, with S\'ersic $n = 2.9$
and $\re \approx 1.6$ kpc.

Clearly, the simplistic, two-component B/D decomposition that is still
standard for large surveys fails dramatically for galaxies like these.
It significantly (even catastrophically) overestimates the fraction of
galaxy light that is part of the ``bulge'', especially when that
component is assumed to be a classical spheroid. Even the somewhat
ad-hoc use of S\'ersic indices to discriminate between classical bulges
and pseudobulges would fail, since for both galaxies the S\'ersic
indices are $> 2$. (These galaxies would satisfy at least two of the
criteria for classical bulges in \citealt{kormendy16}, since they have
$n >2$ \textit{and} $B/T \ga 0.5$.) This is misleading because both
galaxies have significant pseudobulges: the B/P bulge in NGC~4608 and
the B/P bulge \textit{and} the nuclear disc in NGC~4643.

We have argued that NGC~4608 has a classical bulge that is only $\sim
12$\% of the galaxy light, while NGC~4643 has (at best) a compact
classical bulge that is $\sim 0.5$\% of the total light; this
latter component could also be interpreted as a large NSC. This means that
two-component fits would overestimate the spheroid fraction of NGC~4608
by roughly a factor of four, and by roughly a factor of at least \textit{one
hundred} for NGC~4643.

It is now (somewhat) common to include optional bars in decompositions
of moderately large samples
\citep[e.g.,][]{gadotti09,salo15,mendez-abreu17,kruk18}, though this is
still not true of the largest surveys
\citep[e.g.,][]{simard11,mendel14,kim16a,lange16,bottrell19}. This does
tend to reduce the fraction of light assigned to the ``classical
bulge'', but can still significantly overestimate things.

In the case of NGC~4608, the ``spheroid/T'' value drops from the
two-component value of 0.49 to 0.33 when a separate (single-S\'ersic)
bar is included \citep{gadotti08}. For NGC~4643, the value drops from
$\sim 0.65$ (various two-component fits) to 0.25 in the three-component
fit of \citet{salo15}. So three-component decompositions, with a
simple S\'ersic or Ferrers component for the bar, still result in
overestimating the spheroid fraction by factors of three (NGC~4608) or
fifty (NGC~4643).

\section{Summary}\label{sec:summary}

We have presented a detailed morphological and stellar-kinematic analysis
of two similar massive, early-type, barred galaxies, NGC 4608 and NGC
4643. We find that images of both galaxies can be fit using a new two-component
bar model which assigns $\sim 14$--17\% of the total galaxy light to the
narrow, vertically thin outer part of the bar and $\sim 30$\% of the
total light to the bar's B/P bulge (the vertically thickened inner part
of the bar). A further $\sim 39$--41\% of the light can be assigned to
the main disc of each galaxy, leaving $\sim 12$--13\% of the light for
the central stellar components inside the bars.

In NGC~4608, this central component is apparently a classical bulge,
with S\'ersic $n = 2.2$ and half-light radius $\re \approx 310$ pc;
there is also evidence for a possible nuclear star cluster, though this
is only $\sim 0.06$\% of the galaxy light. Published SAURON stellar
kinematics show slow rotation and high velocity dispersion in this
region, consistent with the idea that this is a classical bulge.

By contrast, the inner light in NGC~4643 is dominated by a stellar
nuclear disc, with a somewhat unusual broken-exponential
surface-brightness profile; the break in this profile produces a
ringlike feature in unsharp masks, though there seems to be little in
the way of an actual ring in addition to the nuclear disc. There is also
evidence for a compact, round structure at the very center, with S\'ersic
$n \approx 1.6$ and $\re \approx 35$ pc, amounting to $\sim 0.5$\% of
the light; this is probably either a compact classical bulge 
\citep[such as some of those identified in][]{erwin15a} or a
very large nuclear star cluster.

The stellar kinematics for NGC~4643, including both published SAURON and
MUSE data (and our reanalysis of the latter), show strong rotation,
lower velocity dispersion, strong $V$-\hthree{} anticorrelation, and
elevated \hfour{} in the inner region, consistent with a rapidly
rotating, kinematically cool nuclear disc. The increased \hfour{} is, we
suggest, due to the overlapping contributions from the (more slowly
rotating, kinematically hot) B/P bulge and the (faster-rotating,
kinematically cool) nuclear disc. The inner $r < 50$ pc shows a distinct
\textit{drop} in the velocity dispersion, in approximately the same
region as where the central compact classical bulge (or large NSC)
dominates. Although there are a handful of other examples of similar
nuclear dispersion drops in some other early-type, barred galaxies, the
explanation for this remains unclear.

The dramatically different central structures in these two otherwise
very similar galaxies indicates that seemingly identical galaxies
can have significantly different centers and thus different formation
histories.

We also note that both galaxies have quite large $B/T$ ratios when they
are fit using the same two-component (bulge + disc) models as are used
for large galaxy surveys. If these galaxies were included in such
surveys, they would most likely be assigned $B/T$ values of $\sim
0.5$--0.7, even though their actual spheroid fractions are only 0.12 for
NGC~4608 and 0.005 for NGC~4643; the high S\'ersic indices for the
``bulges'' in such 2-component fits ($n \sim 2.5$--3.5) would
(erroneously) suggest the existence of dominant classical bulges in
these galaxies. Even three-component fits (disc + S\'ersic/Ferrers bar +
bulge) produce $B/T$ values of 0.25--0.3, which significantly
overestimates the spheroid fraction.

This paper is a preview of the analysis we plan to apply to the
other $\sim 50$ galaxies in our volume- and mass-limited sample of
nearby S0 and early/intermediate-type spirals, as part of a project
to determine the true nature and characteristics of the central
stellar structures of disc galaxies.

\section*{Acknowledgments} 

We thank the referee for their comments on this paper, which have led to
several improvements. We are also happy to thank Dimitri Gadotti,
Matias Bla{\~n}a Diaz and Mattia Sormani for comments on earlier drafts of this
paper. VPD was supported by Science and Technology Facilities Council
Consolidated grant ST/R000786/1. AdLC acknowledges support from grant
AYA2016-77237-C3-1-P and JMA acknowledges support from grant
AYA2017-83204-P, both from the Spanish Ministry of Economy and
Competitiveness (MINECO).

This research is based on observations made with the NASA/ESA
\textit{Hubble Space Telescope}, obtained at the Space Telescope Science
Institute. Support for Program number GO-15133 was provided by NASA
through a grant from the Space Telescope Science Institute. The Space
Telescope Science Institute is operated by the Association of
Universities for Research in Astronomy, Incorporated, under NASA
contract NAS5-26555.

This work is based in part on observations made with the
\textit{Spitzer} Space Telescope, obtained from the NASA/IPAC Infrared
Science Archive, both of which are operated by the Jet Propulsion
Laboratory, California Institute of Technology under a contract with the
National Aeronautics and Space Administration. This paper also makes use of
data obtained from the Isaac Newton Group Archive which is maintained as
part of the CASU Astronomical Data Centre at the Institute of Astronomy,
Cambridge.

Based on observations collected at the European Organisation for
Astronomical Research in the Southern Hemisphere under ESO programme
097.B-0640(A).

Funding for the creation and distribution of the SDSS
Archive has been provided by the Alfred P. Sloan Foundation, the
Participating Institutions, the National Aeronautics and Space
Administration, the National Science Foundation, the U.S. Department of
Energy, the Japanese Monbukagakusho, and the Max Planck Society. The
SDSS Web site is \texttt{http://www.sdss.org/}.

The SDSS is managed by the Astrophysical Research Consortium (ARC) for
the Participating Institutions.  The Participating Institutions are
The University of Chicago, Fermilab, the Institute for Advanced Study,
the Japan Participation Group, The Johns Hopkins University, the
Korean Scientist Group, Los Alamos National Laboratory, the
Max-Planck-Institute for Astronomy (MPIA), the Max-Planck-Institute
for Astrophysics (MPA), New Mexico State University, University of
Pittsburgh, University of Portsmouth, Princeton University, the United
States Naval Observatory, and the University of Washington.

This research also made use of both the
NASA/IPAC Extragalactic Database (NED) which is operated by the Jet
Propulsion Laboratory, California Institute of Technology, under
contract with the National Aeronautics and Space Administration, and the
Lyon-Meudon Extragalactic Database (LEDA; http://leda.univ-lyon1.fr).

\section*{Data Availability}

The raw \textit{HST}, \textit{Spitzer}, and MUSE data for both galaxies
are publicly available from the respective archives. Reduced, mosaic
\textit{Spitzer} images for individual observations of both galaxies are
also available from the \textit{Spitzer} archive, and the reduced
\sfourg{} mosaic image for NGC 4643 is available at the NASA/IPAC
Infrared Science Archive
(\url{https://irsa.ipac.caltech.edu/data/SPITZER/S4G/}). The reduced,
combined MUSE datacube for NGC 4643 is available at
\url{http://archive.eso.org/wdb/wdb/adp/phase3_spectral/form?
collection_name= MUSE_DEEP}; our kinematic analyses of this datacube are
available on request. Finally, the SAURON kinematic data for both
galaxies is available at the ATLAS3D site
(\url{http://www-astro.physics.ox.ac.uk/atlas3d/}.

Reduced images of our WFC3-IR F160W observations of both galaxies, along
with sky-subtracted versions of the IRAC1 images, mask and PSF
images, ellipse-fit files, \Imfit{} config files, and Python code for
generating the figures in this paper can all be found in a Github
repository at \url{https://github.com/perwin/n4608-n4643} and also at
\url{https://doi.org/10.5281/zenodo.4235501}.

\bibliographystyle{mnras}

\appendix{}

\section{The Broken-Exponential Nuclear Disc in NGC~4643}\label{app:nd}

For NGC~4643, we argue that the inner region is best understood as a
nuclear disc with a somewhat unusual broken-exponential radial
surface-brightness profile. The evidence for a distinct nuclear disc of
\textit{some} kind is relatively unambiguous. First, there is set of
isophotes (semi-major axes $\sim 1$--4.5\arcsec) with similar
ellipticities and, most significantly, a common position angle ($\sim
52\degr$) which is clearly distinct from that of the bar ($\sim
133\degr$); see, e.g., Figures~\ref{fig:n4643-overview} and
\ref{fig:efits}. Second, unsharp masking shows a clear, narrow
``ring-like'' zone at $r \sim 3\arcsec$ where the surface-brightness
changes abruptly (e.g., Figure~\ref{fig:n4643-umasks}). And, of course,
there is very clear stellar-kinematic evidence for a kinematically cool
and thus disclike structure in this region
(Section~\ref{sec:kinematics}).

\begin{figure}
\begin{center}
\hspace*{-3mm}\includegraphics[scale=0.9]{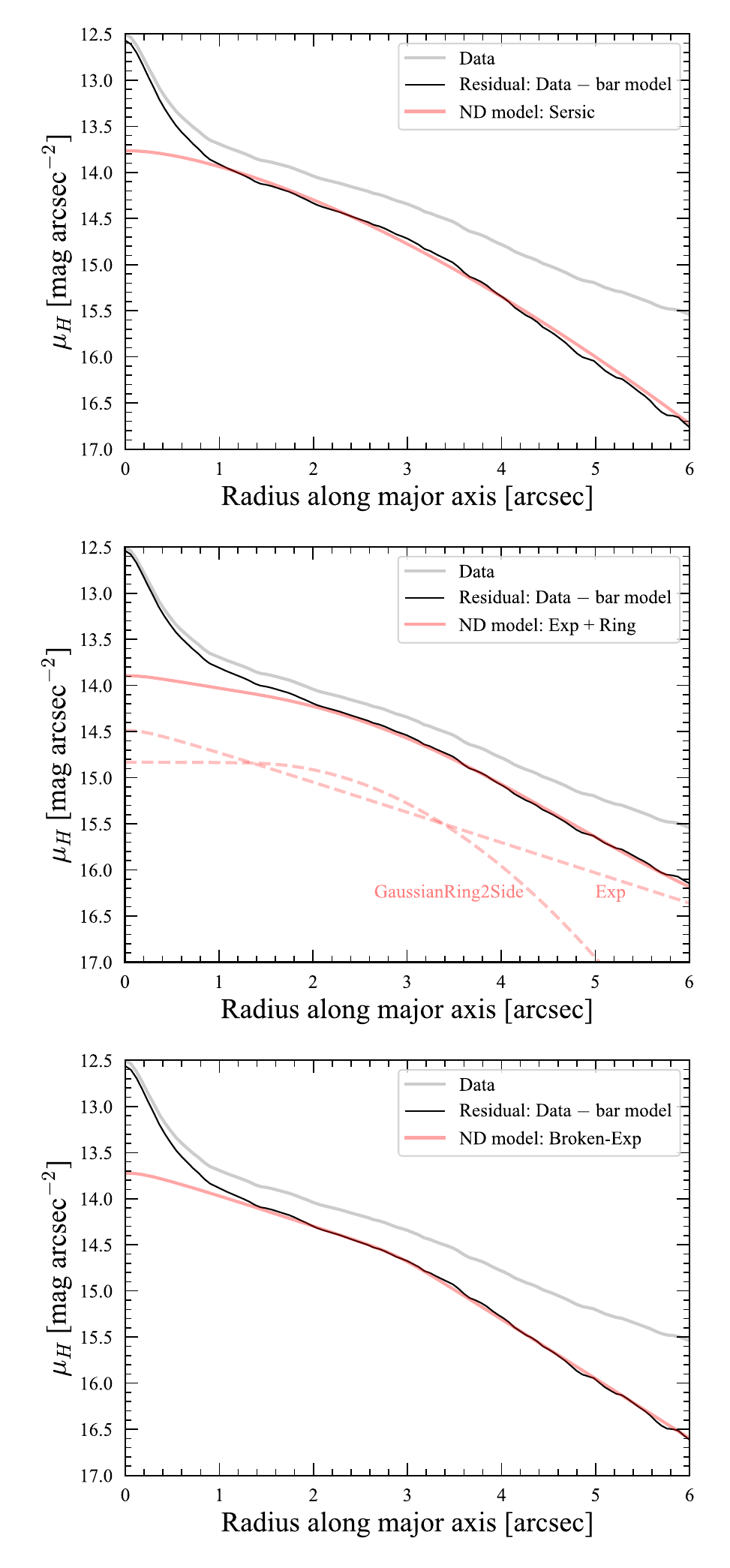}
\end{center}

\caption{Profiles from major-axis cuts through the F160W image of NGC~4643
(thick grey lines) and through residual images (black lines), formed
by subtracting the bar components of the best fitting model.
All models are identical except for how the nuclear-disc region is 
represented. In all three cases, the residuals in the nuclear-disc
region show similar broken-exponential profiles with a break at $r \sim 3\arcsec$,
indicating that this is a robust feature of the galaxy.
\textbf{Upper panel:} Best-fitting model uses S\'ersic component for
nuclear disc. \textbf{Middle panel:} Best-fitting model uses Exponential
+ GaussianRing2Side components for nuclear disc. \textbf{Lower panel:}
Best-fitting model uses BrokenExponential component for nuclear disc
(this is the model discussed in the main text of the paper).}\label{fig:n4643-nd-profiles}

\end{figure}

However, the exact profile of this structure is perhaps not as clear.
For example, might the ringlike feature in the unsharp mask be the
result of a narrow ring added to a smoother profile (e.g., that of an
exponential)? Figure~\ref{fig:n4643-nd-profiles} plots major-axis
profiles (folded about $r = 0$) of residual images obtained by
subtracting the bar+main-disc components of three different best-fit
models. The goal is to see how much variation in the residual profile is
due to variations in the total model. All three models use our standard
composite S\'ersic + FlatBar model for the bar and a small S\'ersic
component for the central classical bulge/NSC; they differ in how we
model the nuclear-disc region itself. The top panel shows the profile
when we use a S\'ersic function for the nuclear disc. The middle panel
shows the same when the model uses the sum of an exponential and an
asymmetric Gaussian ring (\Imfit's GaussianRing2Side component) -- a
plausible model for the sum of a star-forming nuclear ring and an
underlying nuclear disc. Finally, the bottom panel shows the result for
our preferred best-fit model (nuclear disc modeled as
BrokenExponential), and is the same as what is shown in the right-hand
panels of Figure~\ref{fig:central-components}.

In all three cases, we see the \textit{same} basic structure in the
residual profile: the compact central excess (which we associate with
the CB/NSC) and a broken-exponential profile outside, which dominates
the residual profile for $r \ga 1\arcsec$. This indicates that our
identification of a broken-exponential structure for the nuclear disc in
NGC 4643 is robust: it persists regardless of the specific components
used in the modeling of this region, and is not an artifact of simply
assuming the profile must be modeled with a BrokenExponential component.

It is also worth noting that in the Exponential + GaussianRing2Side
model, the best-fit ``ring'' radius is only 1.22\arcsec{}, which is
clearly much smaller than the visible break radius ($r \sim 3\arcsec$);
in addition, the inner part of this component is essentially flat, so it
is not really a ring. Although this more complicated model provides a
formally better fit with lower AIC than the BrokenExponential model, its
parameters are less useful as descriptions of the structure. Since the
best-fit BrokenExponential break radius of 2.9\arcsec{} is a much closer
match to the visible break, we prefer it as an approximation of the
nuclear disc.

\section{A Two-Dimensional Model of a Bar with a Boxy/Peanut-shaped Bulge}\label{app:bar}

Although it is standard to model bars in 2D decompositions as simple
elliptical\footnote{With the option of describing the isophotes as
having fixed deviations from pure ellipticity in the discy/boxy sense.}
structures with either S\'ersic profiles or variations on the
\citet{ferrers1877} ellipsoid, recent analysis of bars in simulations
and real galaxies \citep[e.g.,][]{laurikainen14,athanassoula15} has
suggested that many bars are better described as \textit{two-component}
systems. In this scheme, one combines an outer, narrow component with a
shallow (sometimes constant) radial surface-brightness profile and an
inner, rounder component with a steeper profile. These can be readily
associated with the dual three-dimensional structure of bars with B/P
bulges: the outer, vertically thin part of the bar has (when seen
face-on) narrow isophotes and a shallow surface-density profile, while
the vertically thick B/P bulge has rounder isophotes and a steeper
surface-density profile.

\citet{laurikainen14} and \citet{athanassoula15} modeled this composite
structure using a slightly modified version of the Ferrers ellipsoid for
the outer part of the bar and a mildly elliptical S\'ersic component,
typically with S\'ersic index $n$ close to 1, for the B/P bulge (the
``barlens'' in their terminology). While this is a clear improvement on
past models for bars, it is only a weak approximation of the outer-bar
structure. Our model attempts to capture three characteristics of the
outer parts of bars seen at low inclinations: \begin{enumerate} \item
The narrow isophote shape of the outer part of the bar;

\item The fact that the surface-brightness profile along the bar's major axis
has a broken-exponential nature, with a shallow (or even constant) profile
breaking sharply to a steeper, quasi-exponential falloff near the end of the bar;

\item The transition between discy isophotes for most of the outer part of the
bar and rectangular isophotes near and at the end of the bar.
\end{enumerate}

We achieve this by creating a model for the outer part of the bar
(``FlatBar'') with intrinsically elliptical isophotes and a
surface-brightness profile which has a broken-exponential form
\citep{erwin08,erwin15b}. This is then modified by interpolating the
profile as a function of angle relative to the bar major axis. Along the
major axis, we have the default broken-exponential profile, with inner
scale length $h_{1}$ and outer scale length $h_{2} < h_{1}$, and with
the transition region specified by the break radius \rbrk{} and the
sharpness parameter $\alpha$. As we move away from the major axis --
i.e., considering radial profiles with angles relative to the major axis
$\dpa > 0$ -- we gradually increase the value of $h_{2}$ using parabolic
interpolation. For $\dpa \geq \dpamax$, $h_{2} = h_{1}$ and the profile
becomes a simple exponential. In addition, we fix \rbrk{} so that it is
constant regardless of \dpa, as opposed to the normal pattern of
reducing \rbrk{} to match the ellipticity of the image function away
from the major axis (as is done in \Imfit's BrokenExponential function.)

\begin{figure*}
\begin{center}
\hspace*{-3mm}\includegraphics[scale=0.9]{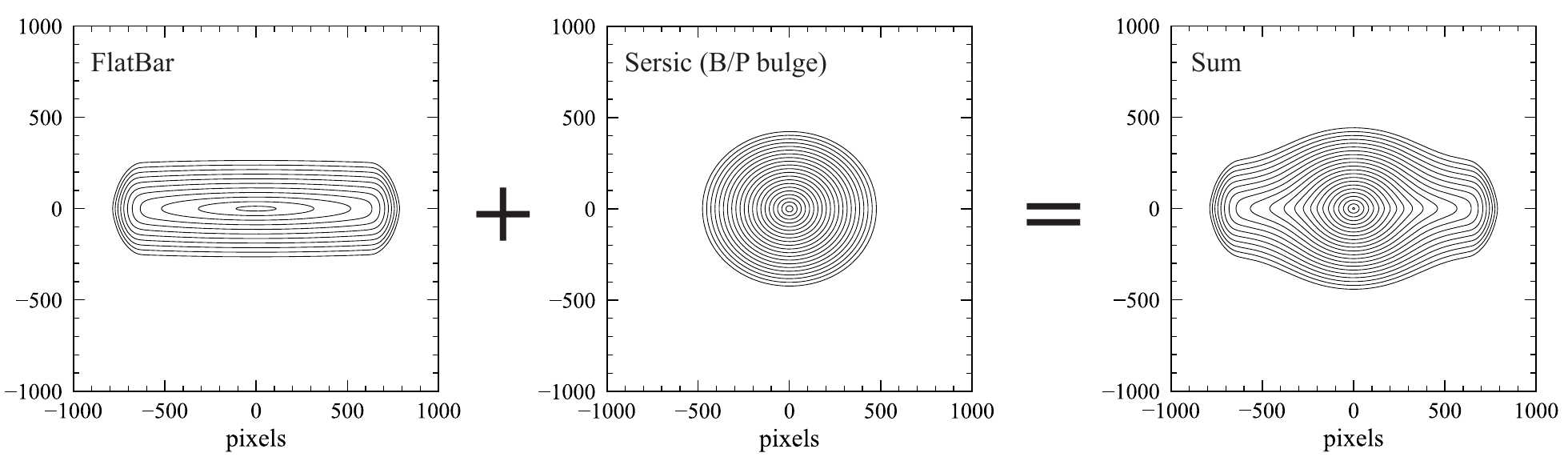}
\end{center}

\caption{The \Imfit{} components of our bar model. Left: the new
``FlatBar'' component, representing a face-on or low-inclination view of
the vertically thin part of the bar. Middle: a mildly elliptical Sersic
component ($n = 1$), representing the face-on/low-inclination view of the B/P
bulge. Right: the sum of the two.  (All isophotes are logarithmically
spaced.)}\label{fig:new-bar-model}

\end{figure*}

As Figure~\ref{fig:new-bar-model} shows, this result in a structure
where the inner isophotes are clearly discy but transition to boxy near
the break radius. At very large radii -- i.e., well outside the bar
proper -- the isophotes can become implausibly polygonal, but this will
usually be in regions dominated by the main disc and so the effects in a
typical combined model will be minimal. 

We combine the FlatBar component with a mildly elliptical S\'ersic
component (for the B/P bulge) to represent the whole bar
(Figure~\ref{fig:new-bar-model}). Preliminary tests for a small number
of barred galaxies indicates that best-fit S\'ersic indices for the B/P
bulge component range between 0.5 and 1.1. For the two galaxies
studied in this paper, we use a Sersic\_GenEllipse component (with
generalized ellipses for the isophotes) to account for the possibility
of slightly boxy (or discy) isophotes in the B/P bulge component.

We emphasize that this is an ad-hoc solution which appears to work well
when the galaxy is close to face-on, and to a certain degree for mildly
inclined systems when the bar is close to either the major or minor axis
-- the latter being the case for the two galaxies studied here. For
intermediate bar position angles and higher inclinations, the projected
appearance of bars with B/P bulges becomes strongly asymmetric \citep[see,
e.g.][]{erwin-debattista13,erwin-debattista17} and this model will be
less suitable.

This function has been added to the most recent release (v1.8) of \Imfit; the code for
it can be found on \Imfit's GitHub site (\url{https://github.com/perwin/imfit}).

\section{A Two-Dimensional Model of a Ring with Azimuthally Varying Intensity}\label{app:ring} 

NGC~4608 has a prominent inner ring surrounding its bar (e.g.,
Figures~\ref{fig:n4608-bestfit-irac1} and
\ref{fig:n4608-bestfit-f160w}). \Imfit{} includes an image function
(GaussianRing) describing an elliptical ring with a Gaussian radial
profile, which we initially used in our modeling of the images of this
galaxy. Although this produced reasonable results, it was clear from both
the images and residuals of the fit that the galaxy's ring did not have
a constant surface brightness, but rather decayed in brightness away
from the major axis of the bar.

To deal with this, we created a modification of the GaussianRing
component in which the peak intensity $A$ of of the (Gaussian) radial profile
varies as a function of angle $\theta$, with $A(\theta) = A_{\rm maj}$
along the ring's major axis ($\theta = 0\degr$), and $A(\theta) = A_{\rm
min}$ at $\theta = 90\degr$ (along the minor axis), interpolating
smoothly between the two values as a function of $\cos(2 \theta)$ for
intermediate angles. The surface brightness at scaled radius $r$ and
angle $\theta$ is thus given by
\begin{equation}
I(r, \theta) \; = \; A(\theta) \exp(-(r - R_{\rm ring})^{2} / \sigma^{2})
\end{equation}
where the ring has semi-major axis $R_{\rm ring}$ and radial width $\sigma$, and
\begin{equation}
A(\theta) = (A_{\rm maj} \, + \, A_{\rm min})/2 + (A_{\rm maj} \, - \, A_{\rm min})/2 \, \cos(2 \theta) .
\end{equation}
(In practice, $A_{\rm min} = A_{\rm min\_rel} A_{\rm maj}$, where $A_{\rm min\_rel}$
is one of the fitted parameters.)

This function -- GaussianRingAz -- has been added to the most recent release (v1.8) of \Imfit; the code for
it can be found on \Imfit's GitHub site (\url{https://github.com/perwin/imfit}).

\section{Parameter Correlations from Bootstrap Resampling}\label{app:bootstrap} 

The parameter uncertainties for our best-fit models for both
galaxies (Tables~\ref{tab:n4608-decomp} and \ref{tab:n4643-decomp}),
based on bootstrap resampling, present a slightly misleading picture,
because they suggest the uncertainties for individual parameters are
independent of each other. In reality, the uncertainties may include
correlations between different parameters.

\begin{figure*}
\begin{center}
\hspace*{-10mm}\includegraphics[scale=0.75]{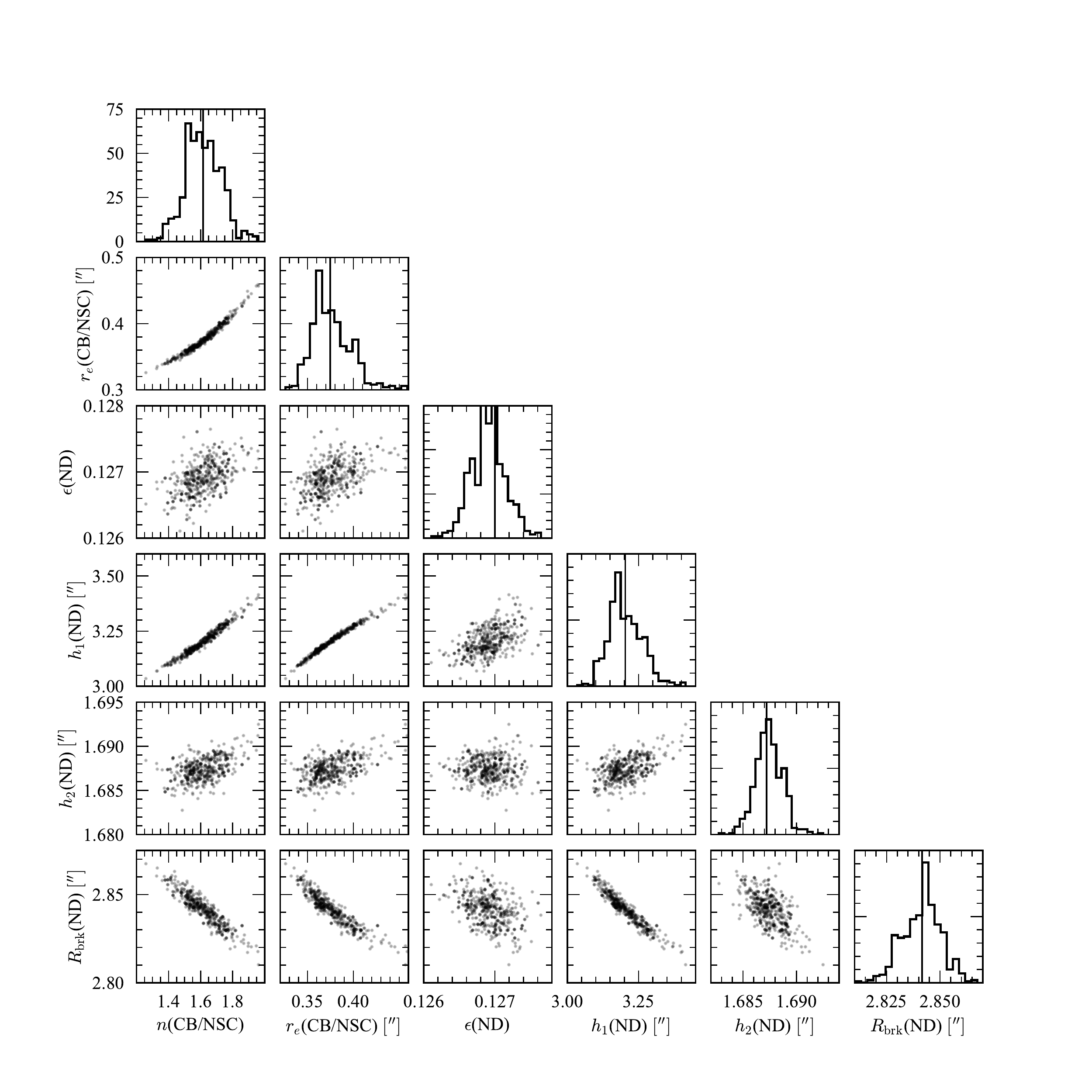}
\end{center}

\caption{Corner plot showing bootstrap resampling analysis for our model
of the F160W image of NGC~4643 (e.g.,
Figure~\ref{fig:n4643-bestfit-f160w}, Table~\ref{tab:n4643-decomp}),
based on 500 rounds of bootstrap resampling. Panels show best-fit
parameter values from fits to the resampled data from each iteration
(grey points) or histograms of parameter values; vertical lines in the
histogram panels show the best-fit parameter values for the fit to the
original data. The full model contains 33 free parameters for five
components; we show a subsample of parameters corresponding to the
S\'ersic component for the classical bulge/nuclear star cluster (``CB/NSC'') and the
BrokenExponential component for the nuclear disc
(``ND'').}\label{fig:bootstrap}

\end{figure*}

In Figure~\ref{fig:bootstrap} we plot parameter distributions from 500
rounds of bootstrap resampling for our fit to the F160W image of
NGC~4643. Since the model has over 30 free parameters, a full corner
plot would need hundreds of individual panels. We show instead a subset
of parameters from the S\'ersic and BrokenExponential components which
fit the innermost structures of NGC~4643: the compact bulge (or NSC) and
the nuclear disc surrounding it. We can see that some parameters are
quite strongly correlated: for example, the S\'ersic index $n$
correlates with the S\'ersic effective radius and with the inner slope
($h_{1}$) of the BrokenExponential, and anti-correlates with the break
radius of the latter. Although this has no effect on our conclusions
in this paper, we note that future analysis of larger samples of galaxies
with similar structures should be careful to ensure that any apparent
correlations between parameters of different components are not simply
side-effects of parameter degeneracies in the modeling.

\bsp	
\label{lastpage}
\end{document}